\documentclass[a4paper,UKenglish]{article}

\usepackage{microtype}
\usepackage{fullpage}
\usepackage{graphics} 

\usepackage{authblk}

\usepackage{graphicx}
\usepackage{amsthm}
\usepackage{amsmath}
\usepackage{enumerate}
\usepackage{color}
\usepackage{xspace}
\usepackage{url}
\usepackage{microtype}
\usepackage{graphicx}
\usepackage{subfigure}
\usepackage{booktabs} 
\usepackage[table,xcdraw]{xcolor}
\usepackage{hyperref}
\usepackage{verbatim}

\graphicspath{{pictures/}}

\usepackage{amsfonts}\usepackage{amssymb}
\usepackage{thmtools}\usepackage{mathscinet}
\usepackage{thm-restate}

    	\definecolor{darkgreen}{rgb}{0.01, 0.93, 0.29}
\definecolor{lightbrown}{rgb}{0.91, 0.4, 0.11}
\usepackage{framed}

\title{BigGraphVis: Leveraging Streaming Algorithms and GPU Acceleration for Visualizing Big  Graphs}
 
\author{Ehsan Moradi}
\author{Debajyoti Mondal} 
 
\affil{Department of Computer Science, University of Saskatchewan, Saskatoon, Canada\\
  \texttt{e.moradi@usask.ca}, \texttt{dmondal@cs.usask.ca}}

\usepackage[textsize=tiny]{todonotes}
\usepackage{verbatim}

\usepackage{microtype}                 
\PassOptionsToPackage{warn}{textcomp}  
\usepackage{textcomp}                  
\usepackage{mathptmx}                  
\usepackage{times}                     
\usepackage{cite}                      
\usepackage{tabu}                      
\usepackage{booktabs}                  
\usepackage{algorithm}
\usepackage{algorithmic}
\usepackage{tabularx,booktabs}
\usepackage{amsmath}
\usepackage{amssymb}
\usepackage{adjustbox}
\usepackage{verbatim}
\begin{document}
\maketitle       
\begin{abstract}
Graph layouts are key to exploring massive graphs. An enormous number of nodes and edges do not allow network analysis software to produce meaningful visualization of the pervasive networks. Long computation time, memory and display limitations encircle the software's ability to explore massive graphs. This paper introduces BigGraphVis, a new parallel graph visualization method that uses GPU parallel processing and community detection algorithm to visualize graph communities. We combine parallelized streaming community detection algorithm and probabilistic data structure to leverage parallel processing of Graphics Processing Unit (GPU). To the best of our knowledge, this is the first attempt to combine the power of streaming algorithms coupled with GPU computing to tackle big graph visualization challenges. Our method 
extracts community information in a few passes on the edge list, and renders the community structures using the ForceAtlas2 algorithm. Our experiment with massive real-life graphs indicates that about 70 to 95 percent speedup can be achieved by visualizing graph communities, and the visualization appears to be meaningful and reliable. The biggest graph that we examined contains above 3 million nodes and 34 million edges, and the layout computation took about five minutes. We also observed that the BigGraphVis coloring strategy can be successfully applied to produce a more informative ForceAtlas2 layout.%
\end{abstract}

\section{Introduction}
{G}{raph} visualization has been one of the most useful tools for studying complex relational data. A  widely used algorithm for computing a graph layout is \emph{force-directed layout}~\cite{kobourov2012spring}, where forces on the nodes and edges are defined in a way such that in an equilibrium state, the distances between pairs of nodes become proportional to their graph-theoretic distances. As a consequence, such layouts can reveal dense subgraphs in a graph. Since sequential computation for big graphs (with millions of nodes and edges) becomes slow, 
there have been several attempts to visualize a small supergraph~\cite{walshaw2000multilevel,hachul2004drawing,DBLP:conf/gd/NachmansonPLRHC15,riondato2017graph}, leverage GPU-computing~\cite{brinkmann2017exploiting},  computing graph thumbnails~\cite{DBLP:journals/tvcg/YoghourdjianDKM18}, or to retrieve a precomputed  visualization based on machine learning approach~\cite{DBLP:journals/tvcg/KwonCM18}.

\begin{figure*}[pt]
  \centering
 \includegraphics[width=\linewidth]{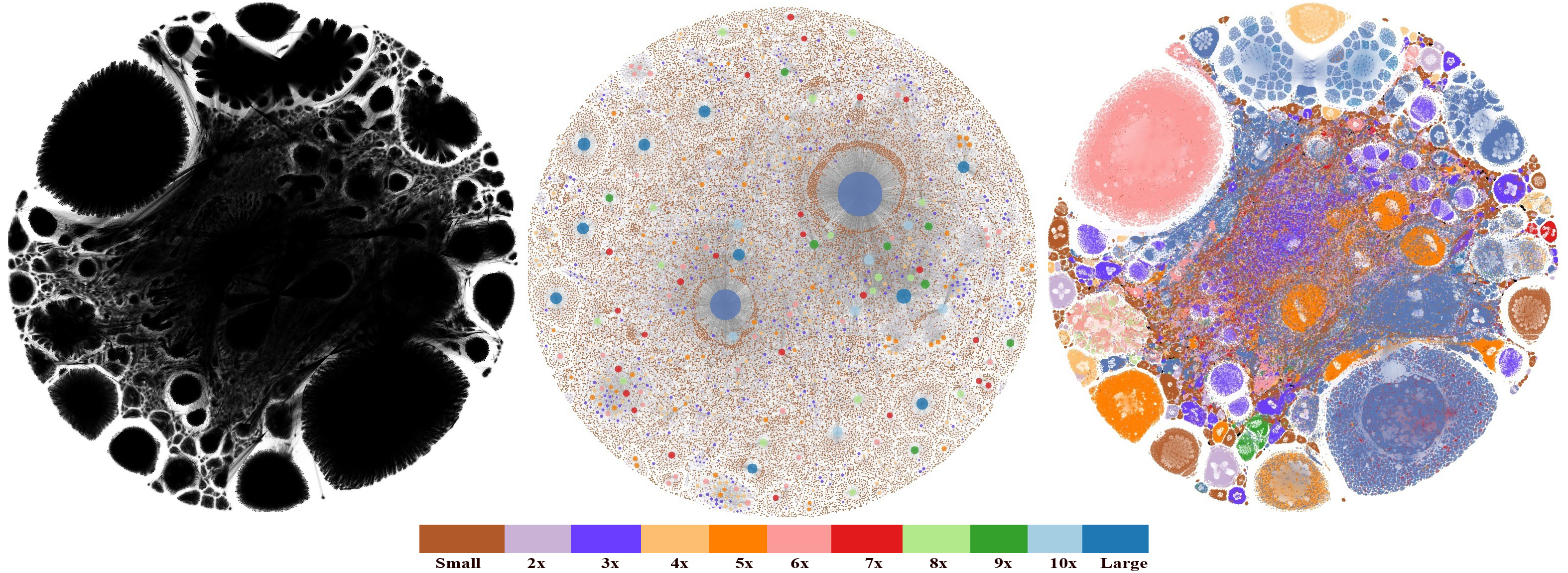}
 \caption{Different layouts computed for a graph (web-BerkStan \cite{leskovec2009community}): (left) A layout produced by the traditional ForceAtlas2 algorithm. (middle) The graph layout produced by our approach --- BigGraphVis, where the communities are shown with colored nodes of varying radii based on the community sizes. (right) The ForceAtlas2 layout, where nodes are colored by the color assignment computed through BigGraphVis community detection. A qualitative color scale is used to provide an idea of the node size distribution and to illustrate a detailed mapping of the nodes between BigGraphVis and ForceAtlas2 visualizations. 
 \label{teaser}
 }
\end{figure*}

In this paper, we consider visualizing a supergraph using parallel computing  that not only retains important information of the original graph but also increases  parallelization capacity. 
 Such a compressed graph can be constructed in various ways. For example, one can examine a sampled or filtered graph~\cite{leskovec2006sampling}, or examine only the densest subgraph~\cite{gallo1989fast} (a vertex-induced subgraph with the maximum average degree), or several dense subgraphs (communities) of the original graph~\cite{gibson2005discovering,newman2004fast}. Sometimes nodes and edges are aggregated~\cite{elmqvist2008zame,balzer2005voronoi} (merged into some supernodes) based on node clusters or attributes. It is thus desirable to have a visualization of a meaningful sample of the graph when visualizing a big graph~\cite{von2011visual} but having it all in a feasible speed. Since supergraph retains major relational structures of the original graph, the hope is that the visualization would potentially reveal the crucial relations between communities. To  achieve parallelism, a common approach is to assign each node/supernode to a processing unit, where the key is to synchronize the outputs in a way such that the  overall processing remains fast and reliably accurate~\cite{moradi2015fast}.

Most algorithms for computing a supergraph runs in linear time on the number of vertices and edges of the graph. However, even linear time algorithms turns out to be slow  for big graphs if the constant factor  hidden in the asymptotic notation is large. A bold idea in such a scenario is to design a \emph{parallel streaming algorithm} that processes the edge list in one pass, takes only a few operations to process each edge, and renders the graph as soon as it finishes reading the edge list. Note that the graph does not necessarily need to be dynamic or streamed, one can read the graph from external or local memory. The idea of a parallel streaming algorithm is to limit the number of passes the input elements are being looked at. Furthermore, to achieve high computational speed, realizing such an approach for big graphs would require streaming edges in parallel. Such streaming algorithms naturally comes with several benefits, e.g., fast computing and limited memory, and with a cost of loosing quality. However, to the best of our knowledge, no such big parallel graph visualization algorithm is known to date.

Understanding whether big graphs can be effectively visualized in a streaming or one-pass model predominantly relies on our knowledge of whether meaningful structure can be extracted in such a  model. In this paper we take a first step towards achieving the goal by bringing the  streaming community detection into the scene, which allows us to create \emph{an edge-weighted supergraph} by merging the detected communities into single nodes (i.e., supernodes). We then visualize the suppergraph using the known graph layout algorithm. Although visualizing a  supergraph is not a new idea, integrating streaming community detection for big graph visualization is a novel approach and brings several natural and intriguing  questions: How fast can we produce a supergraph visualization using streaming community detection? Is there a reasonable GPU-processing pipeline? Do we lose the quality significantly using streaming community detection, when compared with the traditional force-based visualization algorithms~\cite{jacomy2014forceatlas2}?  Can we meaningfully stylize traditional graph layouts by coloring the communities without increasing the time overhead?

To keep the whole pipeline of  community detection, graph aggregation, and visual rendering meaningful and fast (in a few minutes), we propose BigGraphVis, which is a GPU-accelerated pipeline that seamlessly integrate streaming community detection and visual rendering. The availability of increased number of computer processing units and GPUs with hundreds of cores have inspired researchers to implement force-based visualization algorithms that leverages these computing technologies~\cite{frishman2007multi,brinkmann2017exploiting}. Although GPUs can allow massive parallelization via high number of cores and low-cost task dispatching, they are structured processing units that suit structured data like matrices \cite{shi2018graph,frishman2007multi}. Hence, designing parallel graph processing algorithms often turns out to be challenging, especially for graphs. Integrating streaming community detection and force-based visualization algorithms becomes even more challenging  since the supergraph between these two process needs to be handled with care retaining as much quality as possible. 

\subsection{Our Contribution}

Our contributions are directed towards examining the feasibility of leveraging parallel streaming community detection for computing fast and reliable visualization for big graphs (Figure~\ref{teaser}). 
\begin{itemize}
 \item {BigGraphVis combines the idea of parallel streaming community detection, node aggregation, and GPU computing to compute and visualize supergraphs, which provides a speed-up factor between 70 to 95 for real-life graphs when  compared with GPU-accelerated ForceAtlas2~\cite{brinkmann2017exploiting} that visualizes the entire graph. } 
 
 Although graph aggregation step creates a smaller graph to visualize step, note that we need to maintain extra data structures to detect communities. We show how a faster speed can be achieved by trading off layout resolution (the amount of details to be visualized), which is crucial for big graphs. 
 
 \item {We observe that ForceAtlas2 visualization often fails to provide us with a proper measure of the 
 the relative community sizes. Since the size of a community is determined by complex force simulation, it is hard to understand whether two communities that are visually similar have a similar number of nodes and edges. Even when two clusters are visually similar with respect to the space they occupy, one may contain a large number of nodes but a small number of edges, and the other may contain a larger number of edges but fewer nodes. Furthermore, a community with a high average degree with fewer nodes may take a significantly large space (due to node repulsion) than another community with a large number of nodes but fewer  edges.}
 
 {We show how a supergraph visualization can mitigate this problem. 
 When we employ the community detection algorithm, we maintain an approximate size (number of edges) for each community, and the supernodes are drawn with circles of various radii based on the community sizes. Hence BigGraphVis can provide a better idea of the number of big communities and the relative sizes of those communities. 
 }
 \item {To compute the supergraph, we first detect communities and then merge each of them into a supernode. Since community detection algorithms take a considerable amount of time, we adopted a streaming community detection algorithm (SCoDA~\cite{hollocou2017linear}) that finds the communities in linear time by going over the edges in one pass. To speed up the procedure, we propose a GPU-accelerated version of SCoDA that allows for hierarchical community detection. 
 }
 \item {We use GPU to manage millions of threads that keep the speed of the whole process. However, computing a supergraph using community detection is challenging in a parallel environment since we need to compute each community's size (i.e., count the number of edges). 
 However, counting is an atomic operation, which means that if we want to count the communities' size in parallel, each thread assigned to a community should go through the whole network. Therefore, we use a data structure count-min sketch~\cite{cormode2011approximating} to approximately compute the size of each community in parallel.
}
\end{itemize}

\section{Related Works}
In this section, we briefly review the literature related to our work. 

\subsection{Community Detection}
{
Community detection in a graph is the procedure of uncovering a group of well-connected nodes. 
A rich body of literature examines community detection algorithms~\cite{lancichinetti2009community,blondel2008fast}, where such algorithms are inspired by different notion of community, e.g., graph partitioning\cite{fan2020application}, hierarchical clustering\cite{murtagh2017algorithms}, spectral clustering\cite{ng2001spectral}, modularity optimization~\cite{blondel2008fast,newman2006modularity,moradi2015fast}, 
random walk~\cite{khan2017network}. 
 In spite of the diversity in community detection algorithms, a community often indicates a well-connected induced subgraph, i.e., the number of edges within a community $C$ is significantly larger than the edges that connect $C$ with the rest of the graph. 
 Some other definitions presented in recent years like graph summarizing \cite{liu2018graph} that trying sample the graph or ignore dome key features of the graph. 
}

{
A common idea of community detection is to define a quality measure and then find a partition of the graph (communities) that maximizes the measure. Modularity~\cite{newman2006modularity,blondel2008fast} and Weighted Community Clustering (WCC)~\cite{liu2014weighted} metrics are some widely used quality measures for community. 
The main problem of the optimization-based approaches is the computation cost. Many approximation algorithms~\cite{fortunato2010community} and parallelization~\cite{fazlali2017adaptive,moradi2015fast,riedy2011parallel} have been proposed to reduce the complexity of the community detection. 
However, many community detection algorithms 
have a strong resistance against  parallelization~\cite{moradi2015fast}. One of the potential areas for parallelization is distributing loop iterations among processing units with the least possible data interference, whereas such iterations for most of the quality measures are not independent,  and thus challenging to parallelize.
}

{
Due to the increasing size of graphs and demand for instant results, researchers have recently proposed streaming algorithms 
that process each edge only once and determine the nodes' communities based on some local rules. 
 SCoDA~\cite{hollocou2017linear} is such an algorithm that we use in our work (the details are in Section~\ref{sec2.3}). 
}


There exists a vast literature on \emph{graph summarization}, which is a series of application-specific algorithms designed to transform graphs into more compact representations while preserving structural patterns, query answers, or specific property  distributions~\cite{liu2018graph,bonifati2020graph}. However, we restrict our attention only to the graph communities as it the most prominent structure revealed by the ForceAtlas2 in big graph visualizations.

\subsection{Force-directed Layouts}
{
Force-directed algorithms~\cite{kobourov2012spring} are widely used to create graph layouts. The idea of force-directed algorithms~\cite{eades1984heuristic,fruchterman1991graph} is to define repulsive force between every pair of nodes and attractive force between adjacent nodes. The algorithm then updates the node positions based on the total force acting on each node. The process is repeated over a number of iterations for better convergence. Some force-directed algorithms are based on energy minimization~\cite{kamada1989algorithm,gansner2004graph}, where the energy is defined such that at a minimum energy state, the distances among the nodes become proportional to their graph-theoretic distances in the graph.
}



{
Computation of pairwise forces is a major bottleneck of a force-directed layout. 
A straightforward calculation of forces take $O(n^2)$ time, but based on an approximation algorithm for performing an $n$-body simulation (Barnes–Hut simulation~\cite{barnes1986hierarchical}), one can compute approximate forces in $O(n\log n)$ time. 
ForceAtlas2~\cite{jacomy2014forceatlas2} is a popular force-directed algorithm that leverages several useful techniques such as Barnes-Hut simulation (for fast force calculation), degree-dependent
repulsive force (to reduce visual cluttering), and local and global adaptive temperatures (for better convergence). ForceAtlas2, implemented in a popular graph drawing software Gephi~\cite{bastian2009gephi}, can produce a good layout for graphs with up to 100,000 nodes. 
}


{Multilevel force-directed layouts~\cite{walshaw2000multilevel,hachul2004drawing} further speeds up computation by first `coarsening' the graph (computing a new graph with smaller size) repetitively in multiple levels. The global energy minimization is  performed across these coarse graphs, and then the detailed drawings  are computed for each coarse graph. 
}
{A number of recent approaches to force-directed layout for big graphs includes using distributed architecture~\cite{arleo2016distributed}, dimensionality reduction~\cite{leow2019graphtsne}, and deep-learning based computation~\cite{haleem2019evaluating}.
}

\subsection{GPU-based Algorithms}
{
Nowadays, GPU computing is increasingly being used for visual analytics of big data. For big graphs, GPU computing can speed up the forces calculation speed~\cite{auber2007improved,jeowicz2013visualization}. A number of GPU implementations for calculating the forces has been proposed in the literature. Godiyal et al.~\cite{godiyal2008rapid} showed that FMM (Fast Multiple Method), a hierarchical approximation algorithm for calculating the repulsion force, can be effectively implemented in GPU. Yunis et al.~\cite{yunis2012scalable} presented a more efficient GPU implementation of the FFM. Frishman et al.~\cite{frishman2007multi} used such a GPU-accelerated FMM to compute force-directed layouts.
}

{
In 2016, Peng Mi et al.~\cite{mi2016interactive} showed that GPU-based force-directed layout algorithms are fast and have potential to support real-time user interaction even for large graphs. In addition to approximate force calculation, they used a multilevel force-directed layout approach. Govert et al.~\cite{brinkmann2017exploiting} showed that the popular force-directed layout algorithm, ForceAtlas2, can also be implemented effectively using GPU with an overall speedup factors between 40 and 123, when compared to the CPU implementations.
}

\section{Technical Background}
\begin{figure}[h]
 \centering
 \includegraphics[width=.8\linewidth]{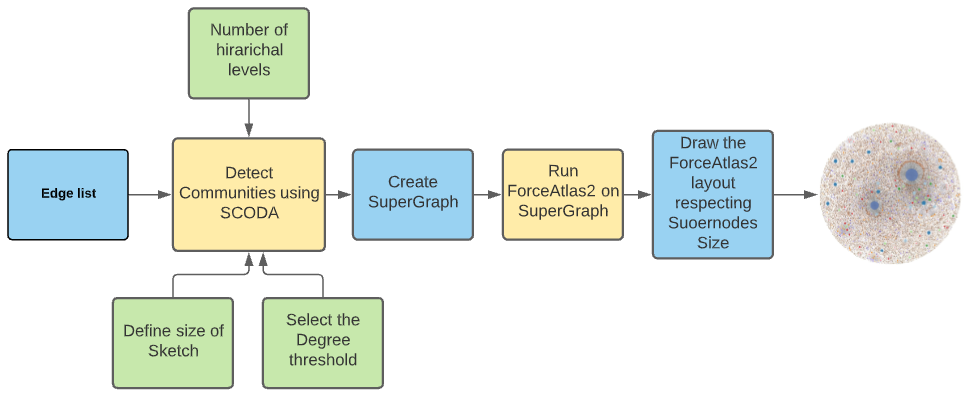}
 \caption{An illustration for the BigGraphViz pipeline.}
 \label{pipeline}
 \end{figure}

{
Our proposed method BigGraphVis combines GPU-accelerated ForceAtlas2, parallel streaming community detection, and count-min sketch data structure (Figure~\ref{pipeline}). Here we briefly describe these components.} 
 
\subsection{ForceAtlas2}
\label{sec:GPUFORCE}
{
We were inspired to choose ForceAtlas2 due to its  capability and popularity~\cite{bastian2009gephi} for producing aesthetic layout for large graphs~\cite{jacomy2014forceatlas2}, and also its speed when implemented via GPU~\cite{brinkmann2017exploiting}. 
The first step of ForceAtlas2 is reading the edge list and putting each node in a random position. After the initialization step, it calculates the following variables.
}
\begin{itemize}
\item {Gravity: The gravity force for each node is towards the center of the drawing space. Hence this force keeps all nodes inside the drawing space. In the GPU-accelerated ForceAtlas2~\cite{jacomy2014forceatlas2}, each thread will manage the gravity force for a group of nodes.}
\item {Attractive Force: This force attracts the neighbor nodes. To avoid conflict among threads, GPU-accelerated ForceAtlas2 distributes edges among the threads. Due to different degrees of nodes, each thread will have a diverse workload, which means they may have a race condition. Atomic operations are used to tackle this problem.}
\item {Body Repulsion: This force moves nodes that are not related further apart from each other. Attractive forces change the position of pairs of nodes that are adjacent but do not affect the nodes that are not, so we need body repulsion to avoid node overlaps. There are different ways to calculate body repulsion. The GPU-accelerated ForceAtlas2~\cite{barnes1986hierarchical} uses an $O(n log n)$-time algorithm based on Barnes–Hut tree (BHT)~\cite{barnes1986hierarchical}. BHT employs a recursive method to create a tree where the nodes represent a hierarchical partition of the space. 
The recursive tree is then used to calculate the repulsive force. GPU ForceAtlas2 calculate BHT based on a GPU implementation of BHT in~\cite{burtscher2011efficient}.}
 \item {Update Speed: This variable controls the node displacements at each iteration. A high speed means the steps are large, which will have less precision; but a low speed means smaller steps and provides better precision. To optimize the convergence, ForceAtlas2 uses swinging. \emph{Swinging} is a strategy to calculate different update speed for different nodes. The most computational cost of update speed calculation is the swinging. Nodes with a high degree need low speed and more precision, but nodes with a small degree need more speed. GPU-accelerated ForceAtlas2 assigns each node a thread to calculate its swinging.} 
\item {Displacement: This variable of the algorithm displaces the nodes based on the forces and the update speed. GPU ForceAtlas2 uses a thread for each node to calculate the new position of each node.
}
\end{itemize}
{
The forces are updated over a number of iterations for better convergence. 
The pseudocode for the ForceAtlas2 is shown in Algorithm~\ref{algo:force}.
}

\floatname{algorithm}{Algorithm}
\renewcommand{\algorithmicfor}{\textbf{For}}
\renewcommand{\algorithmicforall}{\textbf{function}}
\renewcommand{\algorithmicrequire}{\textbf{Input:}}
\newcommand{\COM}[2][.3\linewidth]{%
 \leavevmode\hfill\makebox[#1][l]{//~#2}}

\begin{algorithm}
\caption{ForceAtlas2~\cite{jacomy2014forceatlas2}}
\begin{algorithmic}[1]
\label{algo:force}
\raggedright

\REQUIRE (Undirected graph edge list, iterations, gravitational, repulsion)
\STATE $GlobalSpeed \leftarrow 1$    
\FOR{each node $v_i$}       
\STATE $Pos_i = random(space);$
\STATE $FM_i = 0$  \ \ \ \ \ \ \ \ \ \ \ \ \ \ \ \ \ \ \COM{Net forces matrix}
\STATE $FMTemp_i = 0$  \ \ \ \ \ \ \ \ \ 
\ENDFOR
\FOR{$j = 1 \ldots Max\_Iterations$}
\STATE Build BHT
\FOR{each node $v_i$}
\STATE$FM_i \leftarrow FM_i - Pos_i$ \ \ \ \ \ \ \ \ \ \ \ \ \ \ \ \ \ \ \ \ \ \ \ \ \ \ \ \ \ \ \ \ \ \ \ \ \ \ \ \COM{Gravity}
\STATE$FM_i \leftarrow FM_i + BHT\_force (Pos_i) $ \COM{Repulsion}
\FOR{each neighbour $v_j$ of $v_i$}
\STATE$FM_i \leftarrow FM_i +\frac{P_i - P_j}{|P_i - P_j|}$ \ \ \ \ \ \ \ \ \ \ \ \ \ \ \ \ \ \ \ \ \ \COM{Attraction}
\ENDFOR
\ENDFOR
\STATE {Update $GlobalSpeed$}
\FOR{each node $v_i$}
\STATE $Pos_i \leftarrow local\_Speed(i)\cdot FM_i$ \ \ \ \ \ \ \ \ \ \ \ \ \ \ \ \ \COM{Displacement}
\STATE $FMTemp_i \leftarrow FM_i$
\STATE $FM_i = 0$
\ENDFOR
\ENDFOR
\STATE {\textbf{function }$localSpeed(i)$} \ \ \ \ \ \ \ \ \ \ \ \ \ \ \ \ \COM{For node $v_i$}
\STATE$\textbf{return}\frac{GlobalSpeed}{1 + \sqrt{GlobalSpeed + |FM_i-FMTemp_i|}}$ \COM{consider swing}
\STATE $\textbf{end function}$

\end{algorithmic}
\end{algorithm}
\subsection{SCoDA}
\label{sec2.3}
{
To attain fast speed, BigGraphVis uses a parallel streaming algorithm for community detection. A \emph{streaming algorithm} takes a sequence of edges as input and produces the output by examining them in just one or a few passes. A streaming algorithm is not necessarily for real-time streaming data, but any graph can be read as a list of edges. SCoDA, proposed by Hollocou et al.~\cite{hollocou2017linear}, is a streaming community detection algorithm, which was implemented using a sequential processing. 
}

{
SCoDA is based on the observation that a random edge picked is highly likely to be an \emph{intra-community edge} (i.e., an edge connecting two nodes in the same community) rather than an \emph{inner-community edge} (i.e., an edge between two different communities). Let $e(C,C)$ and $e(C,\overline{C})$ be the intra and inner community edges of a community $C$, respectively. Assume that $e(C) = e(C,C)+e(C,\overline{C})$. Then if we draw $k$ edges from $e(C)$, then the probability that they are all intra-community edges of $C$ is as follows~\cite{hollocou2017linear}.
\begin{equation}
\mathbb{P}[intra_k(C)]=\prod_{l=0}^{k-1}\frac{|e(C,C)|-l}{|e(C)|-l} = \prod_{l=0}^{k-1}(1-\phi_l (C))
\end{equation} 
where $\phi_l (C) = \frac{|e(C,\overline{C})|}{2|e(C,C)|+|e(C,\overline{C})|-l}$. For a well defined community, $\phi_l (C)$ will be small. Therefore, as long as $k$ is small, the chance for picking edges within the community $C$ is large.
}

{
The algorithm starts with all nodes having degree 0 and a degree threshold. It then updates the node degrees as it examines new edges. For every edge, if both its vertices are of degree less than $D$, then the vertex with a smaller degree joins the community of the vertex with a larger degree. Otherwise, the edge is skipped. The degree threshold ensures that only the first few edges of each community are being considered for forming the communities.
}


\subsubsection{Our Modification (GPU-accelerated SCoDA)} 
\label{sec:GPUSCoDA}
{Although SCoDA processes each edge once with two comparisons, we need to deal with graphs with millions of edges. 
Consequently, we design 
a GPU-accelerated version of the SCoDA, where we read the edges in parallel and use atomic operations for the degree update. 
We run SCoDA in multiple rounds such that the communities converge and the number of communities becomes small. This can also be seen as hierarchical community detection. The pseudocode for this process is illustrated in Algorithm~\ref{algo:scoda} (lines 8--22).
 At the end of the first round, some communities are detected, but the number of detected communities is very large. 
Furthermore, the degree of each node is at most the initial degree threshold. In the subsequent rounds, communities with large average degrees absorb the smaller ones. However, due to the increase in node degrees, a bigger threshold is needed. 
Therefore, we update the degree threshold at each round by multiplying it with a constant factor greater than 1. One can choose a factor for the threshold based on the nature of the graph. The choices of such parameters are discussed in Section~\ref{sec:para}. 
}

\subsection{Count-Min Sketch}
\label{sec:count}

{BigGraphVis computes a supergraph based on the detected communities. The communities are represented as `supernodes' with weights corresponding to their number of edges. However, computing frequencies (community sizes) is highly costly for a parallel algorithm (each thread needs to go through the whole data, which is not efficient). The commonly used method will be an atomic operation, which is very time consuming for big graphs. Therefore, we exploit an approximate method, which is reasonable since we are interested in presenting supergraphs and thus can avoid computing finer details. 
A simple solution is to use a hash table to map the data to their occurrences. However, for big graphs, to get a good approximation with this method, one needs to allocate a massive space in the memory. Hence we can use a data structure named count-min sketch~\cite{cormode2011approximating}, which can keep the occurrences a limited space with a better guarantee on solution quality. The Algorithm~\ref{algo:count} describes the count-min sketch used in BigGraphVis. 
}

\begin{algorithm}
\caption{Count-Min Sketch~\cite{cormode2011approximating}}
\begin{algorithmic}[1]
\label{algo:count}
\REQUIRE (A list of events)
\FOR{each event $x$ and hash function $hash_i$}
\STATE $M[i,hash_i(x)] = M[i,hash_i(x)]+1$
\ENDFOR
\STATE{\textbf{return} the minimum value for each column of $M$}
\end{algorithmic}
\end{algorithm}
{
The count-min sketch algorithm maintains a $r\times c$ matrix $M$, where $r, c $ are  determined based on the tolerance for error. 
 Each row is assigned a hash function, and the columns keep an approximate count determined by that hash function. 
 To count the frequency of events, for each event $j$,
 the entry $M[i,hash_i(j)]$ is increased by 1, where $1\le i\le r$ and $hash_i(\cdot)$ is the hash function associated to the $i$th row. The value $min_{1\le i\le r} M[i,hash_i(j)]$ determines the number of occurrences of $j$.
 Having more pairwise independent hash functions ensures less collision and thus provides more accurate results. 
 Since the hash functions are independent of each other, this naturally allows for parallel processing. Altought the count-min sketch basicaly is counter on a matrix but we tried to propose a GPU version of the algorithm which it gave us a better speed but acceptable result since the comparison of results between parallel and sequential shows negligible diffrence.  
 }
\section{Method}
Here we have an overview of BigGraphVis and its parameters.
\subsection{Algorithm Overview}
{The proposed algorithm reads an edge list stream as the input. We then detect the communities based on the GPU-accelerated SCoDA described in Section~\ref{sec:GPUSCoDA}. We then compute the supergraph by representing the communities as supernodes, where each node is weighted proportional to the number of edges it contains. Although one can calculate the communities sizes in the community detection process, that requires adding more atomic counter, which slows down the computation. Hence we leveraged count-min sketch (Section~\ref{sec:count}). More precisely, we took the 
sum of the vertex degrees (equivalently, twice the number of edges) within a community as the weight of the corresponding supernode. To compute this, for each node $v$, we increment $M[i,hash_i(com(v))]$ by the degree of $v$, where $1\le i\le r$, and $hash_i(\cdot)$ is the hash function associated to the $i$th row. 
 The value $min_{1\le i\le r} M[i,hash_i(com(v))]$ determines the approximate size for $com(v)$. 
 Finally, we leveraged the GPU-accelerated ForceAtlas2 (Section~\ref{sec:GPUFORCE}) to draw the aggregated nodes. When drawing a supernode, we choose the radius proportional to the square root of its size. For dense communities, the space occupied by a supernode is thus proportional to the number of vertices that it contains. The coloring of the supernodes is described in Section~\ref{sec:coloring}. 
}

{
 If a visualization for the whole graph is needed (instead of a supergraph), then we first compute a layout for the whole graph using GPU-accelerated ForceAtlas2 and then color the nodes based on the communities detected using SCoDA. The details of such a coloring are described in Section~\ref{sec:coloring}. 
}

\subsection{Parameter Choices}
\label{sec:para}
In BigGraphVis layout, we have some optional parameters which users can define at the beginning of the visualization. 
Here we discuss these parameters. 
\subsubsection{Count-Min Sketch}
{
The error in the count-min sketch can be controlled by choosing the size of the sketch matrix, i.e., the number of hash functions and the number of columns. The fraction that can collide with an item in the matrix is $\frac{N}{w}$, where $N$ is the total number of items. Although this can vary in practice, Cormode and Muthukrishnan~\cite{cormode2011approximating} observed that the probability of seeing a collision of more than an expected amount is bounded by $\frac{1}{2}$, and for $d$ hash functions, the probability of having a large error is bounded by $\frac{1}{2^d}$. 
This indicates that a larger number of hash functions is a better choice when  accuracy is important. However, this also increases the size of the count-min sketch matrix. In our experiment in this paper, we choose the number of hash functions to be four and the number of columns to be a fraction $10^{-4}$ of the number of edges, which is bounded by our available GPU memory. Choosing a larger number of columns can  improve the count-min sketch accuracy further since more collision can be avoided.
}

\subsubsection{Community Detection}
{
For streaming community detection, we need to define two parameters: the degree threshold and the number of rounds. The streaming community detection algorithm is based on the idea that the probability of intra-community edges appearing before the inter-community edges is very high when a community is being formed. Therefore, the degree threshold could be very sensitive since a very small threshold might miss some intra-community edges, which can break communities into sub-communities \cite{hollocou2017linear}. Similarly, if the degree threshold is selected too high, it may lose granularity, i.e., it can merge too many communities into a single community. Thus, as suggested in the original SCoDA~\cite{hollocou2017linear}, we have chosen the 
most common degree (mode degree $\delta$) in the graph as the degree threshold. 
However, if the user wants to have bigger communities with a larger number of nodes, then choosing a slightly bigger degree will produce such results. In our experiment, we observed a few rounds suffice to have maximum modularity. Since after achieving maximum modularity no communities merging in after rounds therefore choosing a big number of rounds wont affect the running time. We choose the $delta^i$ to be the threshold at the $i$th round, where $i$ runs from 1 to 10. After ending each round the degree of the supernodes increased therefore for merging communities we needed to increase the degree threshold.
}

\subsubsection{Convergence}
{The node positions in the ForceAtlas2 algorithm updates in several rounds so that the energy of the system is minimized. To achieve convergence, one needs to choose a large number of iterations for big graphs. For the graphs with millions of nodes and edges, the GPU-accelerated ForceAtlas2 with 500 iterations showed a good performance~\cite{brinkmann2017exploiting}. 
However, in our method, much fewer rounds are enough since we have a much smaller network for drawing after community detection. We observed that for visualizing the supergraph, 100 iterations is more than enough to obtain a stable layout for all graphs we experimented with since no re-position happened after 70 rounds for the biggest graph. 
}
\subsection{Coloring the Communities}
\label{sec:coloring}
{
When visualizing supergraph, we create 11 node groups and color them using a qualitative color scheme~\cite{colorbrewer}. Specifically, we first compute the sum $\alpha$ of the sizes of all communities, then sort the communities based on their sizes, and color the smaller communities that take 50\% of $\alpha$ with a brown color. The rest of the supernodes are partitioned into 10 groups and colored using (from small to big) brown, light purple, purple, light orange, orange, light red, red, light green, green, light blue, blue. Such a coloring provides a sense of the community size distribution in the layout.
Note that the above coloring scheme assigns a color to each supernode. If visualization of the whole graph is needed (instead of a supergraph), then we color the layout of the whole graph computed by the GPU-accelerated ForceLayout2, where each node is drawn with the color of its corresponding supernode. Such a compatible coloring provides us a way to examine the quality of ForceAtlas2 from the perspective of community detection and vice versa.
} 



\floatname{algorithm}{Algorithm}
\renewcommand{\algorithmicfor}{\textbf{EachThread}}
\renewcommand{\algorithmicor}{\textbf{GPUdo}}
\renewcommand{\algorithmicrequire}{\textbf{Input:}}

\begin{algorithm}
\caption{BigGraphVis Layout}
\begin{algorithmic}[1]
\label{algo:scoda}
\REQUIRE Edge list ($src,dst$) and a degree threshold $\delta$
\STATE $Threashold \leftarrow Average Degree$
\STATE $ \# GPUthreads \leftarrow \# edges$
\FOR{$i$}
\STATE $D_i \leftarrow 0$
\STATE $C_i \leftarrow i$ 
\STATE $Csize_i \leftarrow 0$
\ENDFOR
\FOR{$j=1 \ldots Number\_of\_Rounds$ } 
\FOR{$i$}
\IF{$deg(src) \leq D \ \AND \ deg(dst) \leq D$}
\IF{$deg(src)>deg(dst)$}
\STATE$C_{dst} \leftarrow C_{src}$
\STATE $deg(dst)++$
\ELSE
\IF{ $deg(dst) >deg(src)$}
\STATE$C_{src} \leftarrow C_{dst}$
\STATE $deg(src)++$
\ENDIF
\ENDIF
\ENDIF
\ENDFOR
\ENDFOR
\newline
\OR {Compute community sizes $Csize_i$} \newline
\OR $C_ix,C_iy \leftarrow ForceAtlas2(Csrc,Cdst,Csize)$ \newline
\OR $Draw(C_i,C_ix,C_iy,Csize_i)$
\end{algorithmic}
\end{algorithm}

\section{Experiments}
In this section, we evaluate the performance of BigGraphVis. 
\subsection{Specifications}
{For our experiments, we used an Nvidia Tesla k20c with 5GB of VRAM. It is based on the Kepler architecture, which has 2496 CUDA cores. The compiler that we used is a CUDA 11.0.194. The experiments' main goal is (I) to see whether BigGraphVis can reduce the computation time when visualizing the supergraph of massive networks when compared to the GPU-accelerated ForceAtlas2, (II) examine how the parameter choices impact the performances of BigGraphVis, and (III) to understand the similarity and dissimilarity when compared with the ForceAtlas2 output. 
}



{For the comparison with GPU-accelerated ForceAtlas2, we used the implementation of 
Brinkmann et al.~\cite{brinkmann2017exploiting}, which is the fastest known to our knowledge. Brinkmann et al.~\cite{brinkmann2017exploiting} mentioned that their implementation is a direct translation of Gephi's Java implementation of ForceAtlas2. 
We used C++ and CUDA~\cite{kirk2007nvidia} for implementing BigGraphVis. Since the layout of the graph aggregate is computed using GPU-accelerated ForceAtlas2~\cite{brinkmann2017exploiting}, which is a direct translation of the well-known ForceAtlas2 algorithm, we do not discuss the layout aesthetics in this paper. 
However, we visually inspect the quality of the detected communities. Although the implementation of ForceAtlas2 that we are using is the same as that of Brinkmann et al.~\cite{brinkmann2017exploiting}, their output provides a grayscale layout. BigGraphVis leverages the community detection to color the nodes. Furthermore, the node repulsion in BigGraphVis considers the node weights (i.e., supernode sizes), which provides the space needed to draw the supernode. 
For a proper comparison of the speed up, we choose the 
ForceAtlas2 force parameters 
similar to Brinkmann et al.'s work. 
Thus the gravitational and repulsive force parameters remain the same as 1 and 80 for all networks. Although according to Brinkmann et al., tuning these variables do not affect the algorithm's performance. 
}

\begin{table*}
\scriptsize
 \caption{BigGraphVis speed up when compared with GPU-accelerated ForceAtlas2 (time is in milliseconds). DT, SN, SE, M are the degree threshold, super nodes, super edges, and modularity, respectively. SG time is the time to compute supergraph. BGV time is the total time taken by BigGraphVis.}
\label{tb-results}
\begin{adjustbox}{width=1\textwidth}
\begin{tabularx}{\textwidth}{@{}l*{12}{c}c@{}}
\toprule
	Network Name & Nodes & Edges & DT & Sketch Size & SN & SE & FA2 time & BGV Time& SG Time & Speedup & M\\ \hline
\midrule

	Wiki-Talk & 2394384 & 5021410 & 5 & 5000 & 112086 & 122797 & 400949 & 28608&3854 & 92 & 0.64\\ \hline
	bio-mouse-gene & 45101 & 14506195 & 5 & 14500 & 193 & 196 & 50016 & 8937&7941 & 82 & 0.88 \\ \hline
	as-Skitter & 1696414 & 11095298 & 7 & 11000 & 136597 & 300779 & 350141 & 58750&7128 & 83 & 0.55\\ \hline
	web-flickr & 105938 & 2316948 & 43 & 2000 & 1094 & 26170 & 25251 & 3280& 1497& 87&0.61 \\ \hline
	github & 1471422 & 13045696 & 11 & 13000 & 71166 & 91345 & 181538 & 17519&9115 & 90&0.90 \\ \hline
	com-Youtube & 1157827 & 2987624 & 4 & 3000 & 211192 & 232266 & 233915 & 43666& 2198& 81&0.73\\ \hline
	eu-2005 & 333377 & 4676079 & 15 & 4500 & 9181 & 20263 & 52268 & 5145&2827 & 90&0.66 \\ \hline
	web-Google & 916427 & 5105039 & 11 & 5000 & 75443 & 125287 & 131792 & 13863&3415 & 89&0.80 \\ \hline
	web-BerkStan & 685230 & 6649470 & 11 & 6500 & 31213 & 57382 & 138000 & 6565& 4566& 95&0.81\\ \hline
	soc-LiveJournal & 3997962 & 34681189 & 17 & 34500 & 248188 & 566160 & 3862325 & 315072&21344 & 91&0.62 \\ \hline
	Authors & 12463 & 10305446 & 2 & 10000 & 4315 & 1398089 & 146443 & 42541&6382 & 70&0.62 \\ \hline

\bottomrule
\end{tabularx}
\end{adjustbox}
\normalsize
\end{table*}

\begin{table*}
\scriptsize
\caption{Running time and number of communities when using one, two, three and four hash functions.}
\label{tb-hash}

\begin{tabularx}{\textwidth}{@{}l*{9}{c}c@{}}
\toprule
Network name&	1hash time & 1hash SN & 1hash SE & 2hash time & 2hash SN & 2hash SE & 3hash time & 3hash SN & 3hash SE \\ \hline
	\midrule

	Wiki-Talk & 24179 & 91656 & 96723 & 23481 & 91900 & 95411 & 20801 & 92040 & 96954 \\ \hline
	bio-mouse-gene & 8918 & 195 & 194 & 8848 & 194 & 193 & 8530 & 196 & 179 \\ \hline
	as-Skitter & 101694 & 135903 & 249492 & 94153 & 135808 & 235315 & 99673 & 135542 & 272123 \\ \hline
	web-flickr & 3265 & 1097 & 25840 & 3273 & 1093 & 25803 & 3763 & 1089 & 25880 \\ \hline
	github & 71094 & 91396 & 33916 & 71103 & 91517 & 33504 & 33052 & 71069 & 91082 \\ \hline
	com-Youtube & 49551 & 2137333 & 279449 & 77684 & 213243 & 272285 & 87877 & 213077 & 285689 \\ \hline
	eu-2005 & 5744 & 9542 & 20932 & 5952 & 8958 & 19444 & 5134 & 9908 & 23054 \\ \hline
	web-Google & 42618 & 75482 & 125153 & 55369 & 75515 & 125010 & 18974 & 75585 & 125686 \\ \hline
	web-BerkStan & 20708 & 31065 & 56956 & 45370 & 31072 & 57192 & 25351 & 31131 & 56206 \\ \hline
	soc-LiveJournal & 289958 & 248462 & 594563 & 314782 & 248206 & 573076 & 288195 & 248678 & 625434 \\ \hline
	Authors & 47713 & 3195 & 762451 & 73283 & 3388 & 792849 & 67724 & 4133 & 1234993 \\ \hline

\bottomrule
\end{tabularx}
\normalsize
\end{table*}

\begin{table*}
\scriptsize
 \caption{Running time, number of communities and their sizes for two, three and four community detection rounds. }
\label{tb-iteration}
\begin{tabularx}{\textwidth}{@{}l*{10}{c}c@{}}
\toprule
Network Name &	2R time & 2R SN & 2R SE & 3R time & 3R SN & 3R SE & 4R time & 4R SN & 4R SE \\ \hline
	\midrule

	Wiki-Talk & 26444 & 114206 & 124718 & 24665 & 113122 & 119821 & 28608 & 112086 & 122797 \\ \hline
	bio-mouse-gene & 8583 & 191 & 174 & 8883 & 191 & 174 & 8937 & 193 & 196 \\ \hline
	as-Skitter & 63279 & 134365 & 211842 & 65134 & 1696414 & 11095298 & 58750 & 136597 & 300779 \\ \hline
	web-flickr & 3286 & 1045 & 24926 & 3322 & 1042 & 24897 & 3280 & 1094 & 26170 \\ \hline
	github & 18069 & 70384 & 89407 & 17568 & 69809 & 88668 & 17519 & 71166 & 91345 \\ \hline
	com-Youtube & 44074 & 210266 & 232222 & 46118 & 209897 & 237955 & 43666 & 211192 & 232266 \\ \hline
	eu-2005 & 4426 & 6357 & 11095 & 4500 & 6100 & 11375 & 5145 & 9181 & 20263 \\ \hline
	web-Google & 13339 & 71657 & 110462 & 13344 & 71387 & 109187 & 13863 & 75443 & 125287 \\ \hline
	web-BerkStan & 9240 & 30516 & 53149 & 9224 & 30652 & 53568 & 9265 & 31213 & 57382 \\ \hline
	soc-LiveJournal & 287109 & 247516 & 596523 & 303297 & 246873 & 593370 & 315072 & 248188 & 566160 \\ \hline
	Authors & 20638 & 9 & 10 & 21485 & 14 & 26 & 47841 & 4315 & 1398089 \\ \hline
\bottomrule
\end{tabularx}
\normalsize
\end{table*}
\begin{figure*}[pt]
 \centering
 \includegraphics[scale=.08]{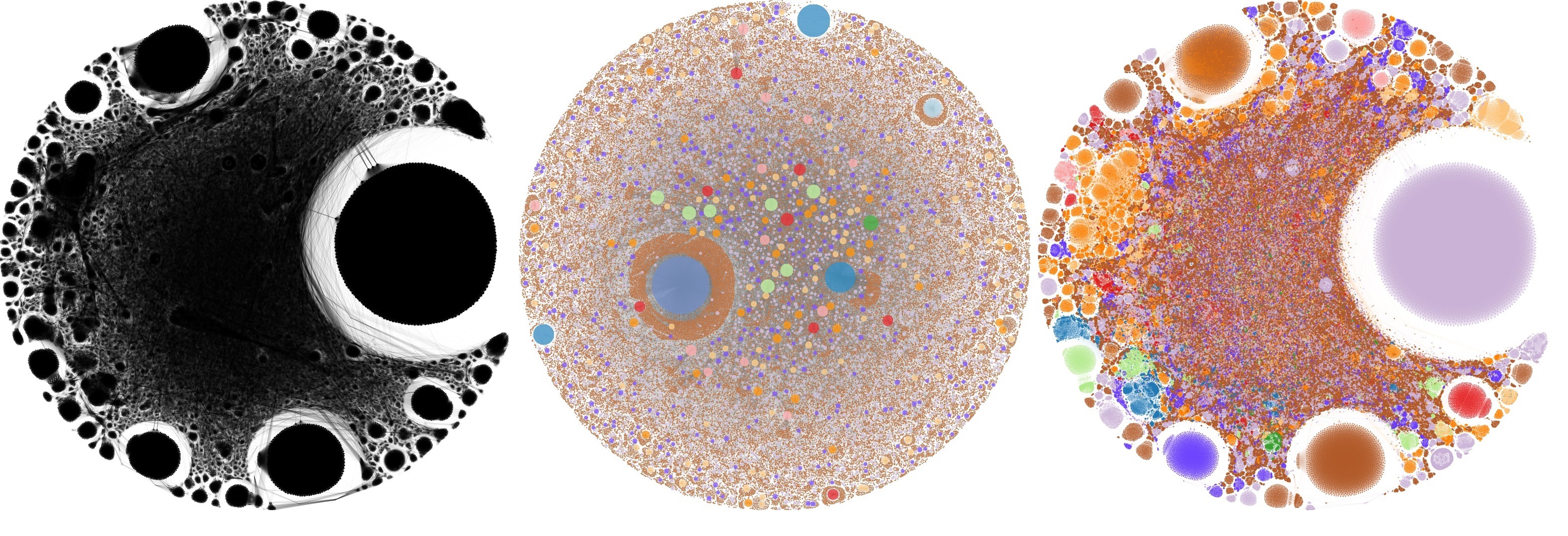}
 \includegraphics[scale=.08]{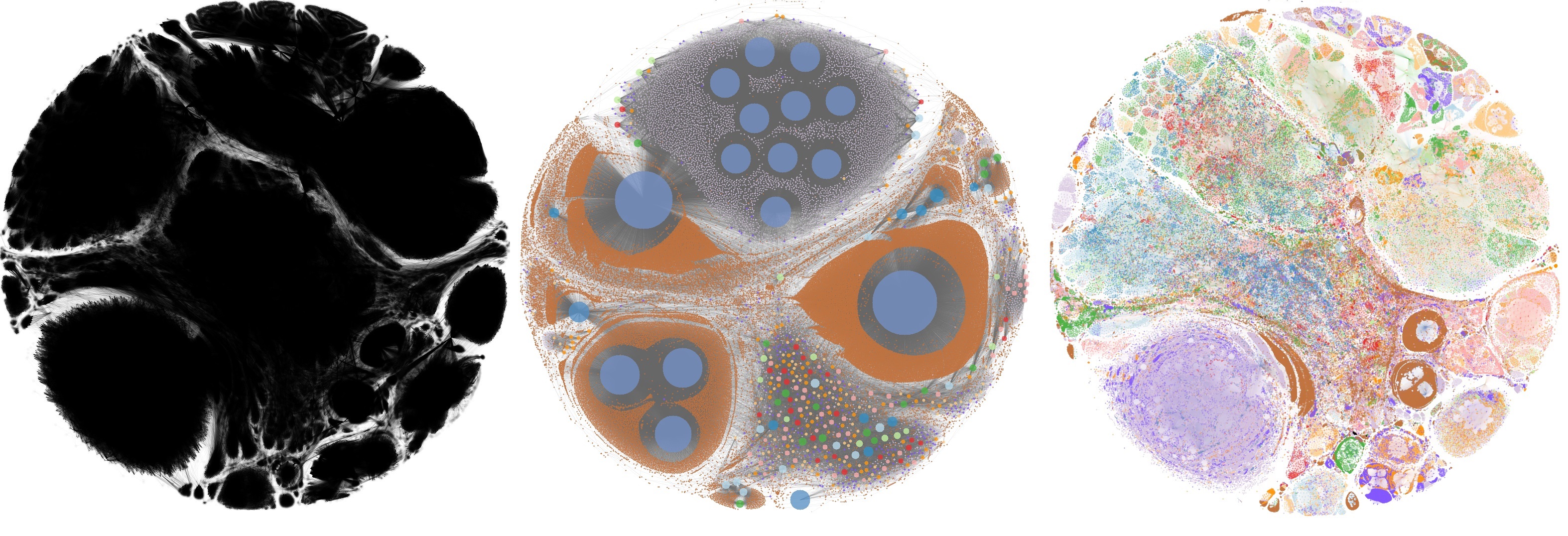}
 \includegraphics[scale=.08]{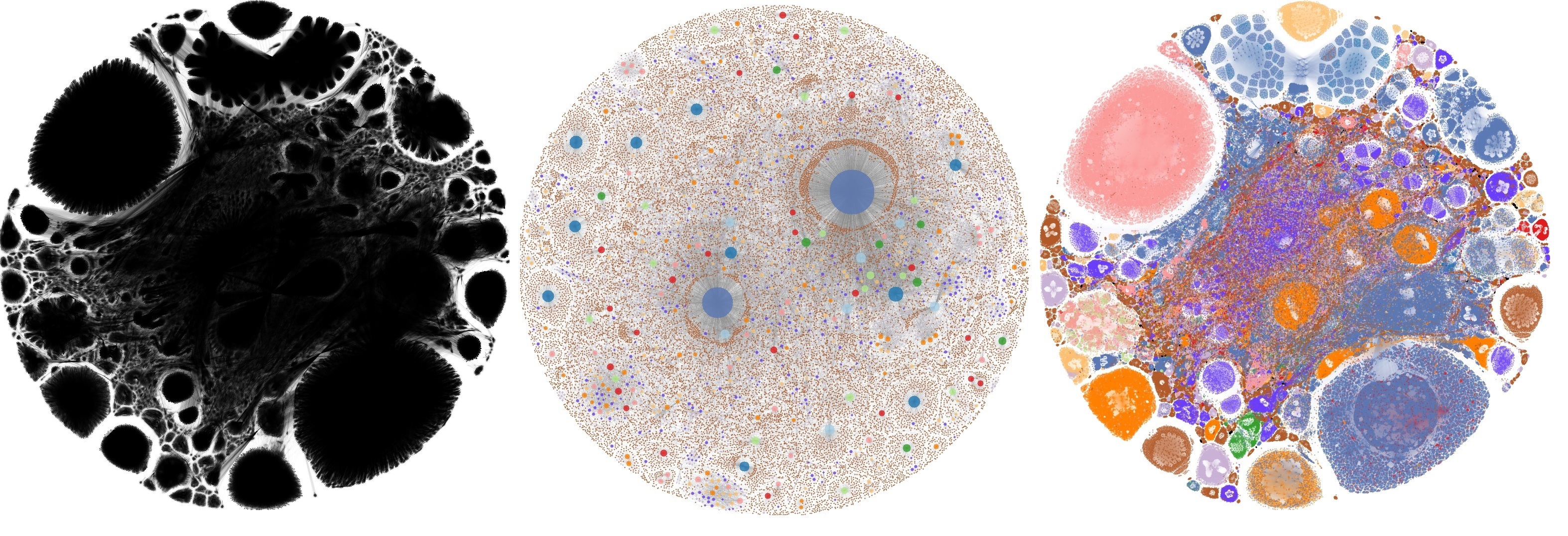}
 \includegraphics[scale=.08]{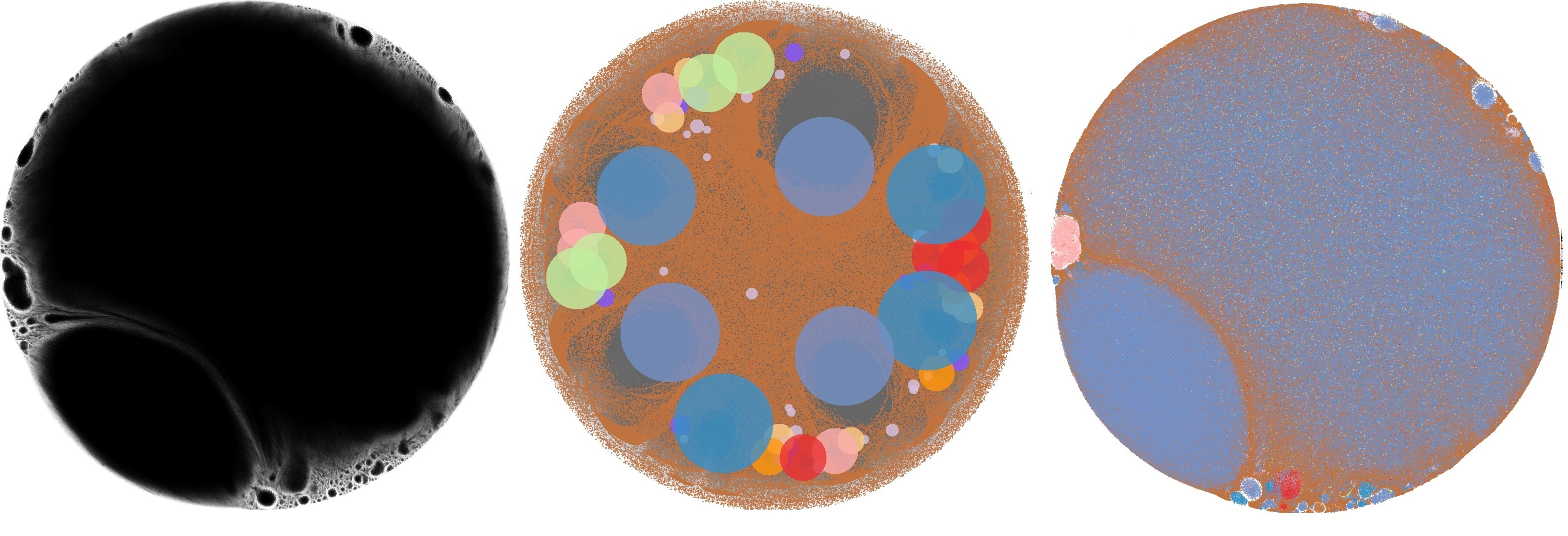}
 \caption{ForceAtlas2 layout, BigGraphVis layout and stylized ForceAtlas2 layout for  four graphs: (top-left) github, (top-right)  eu-2005, (bottom-left) web-BerkStan and (bottom-right) soc-LiveJournal. 
 }
 \label{fig:vis}
\end{figure*}

\begin{figure*}[pt]
 \centering
    \includegraphics[scale=.07]{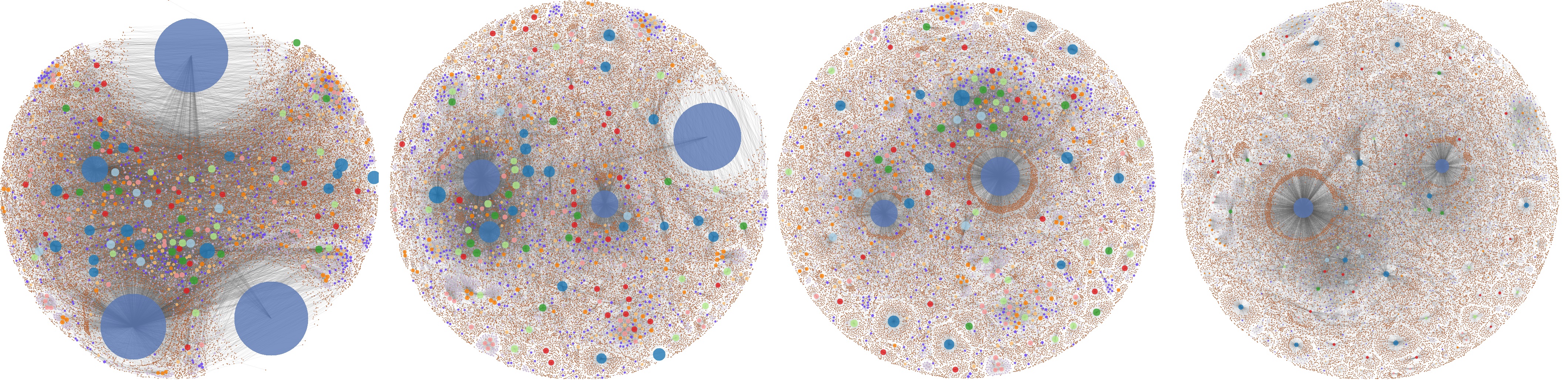}
    \includegraphics[scale=.07]{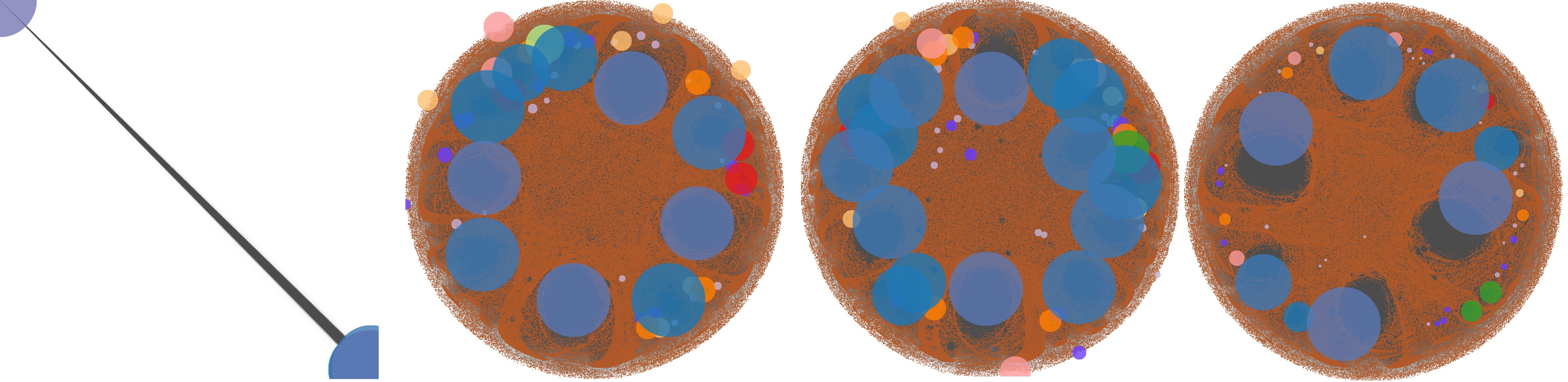}
 \caption{Illustration for the effect of using one to four hash functions for the web-BerkStan (left) and soc-LiveJournal (right). 
 Four images per graph, i.e., from left to right, 1, 2, 3 and 4 hash functions are used to visualize each graph.}
    \label{fig:hash}
\end{figure*}
\begin{figure*}[pt]
 \centering
 \includegraphics[scale=.08]{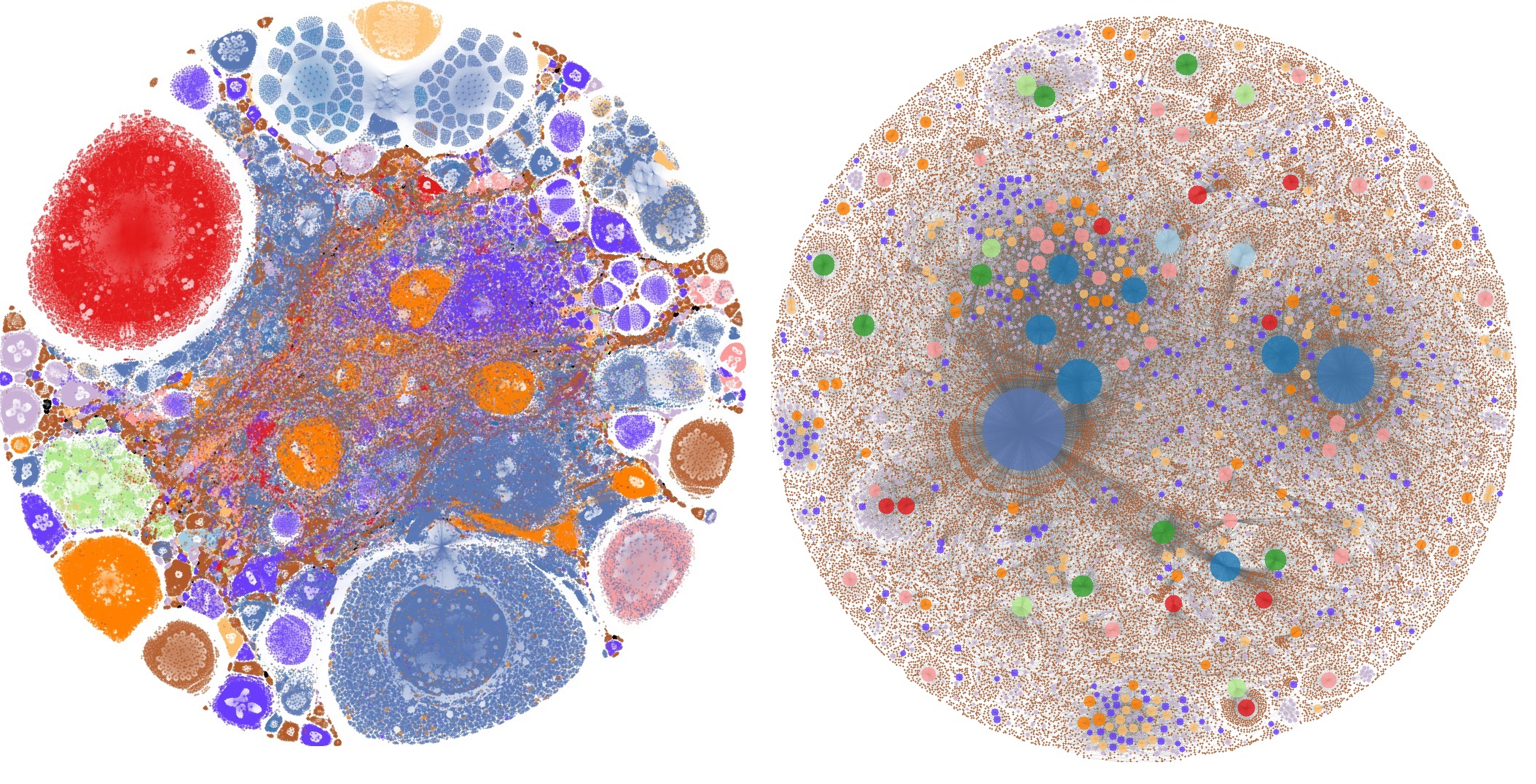}
 \includegraphics[scale=.08]{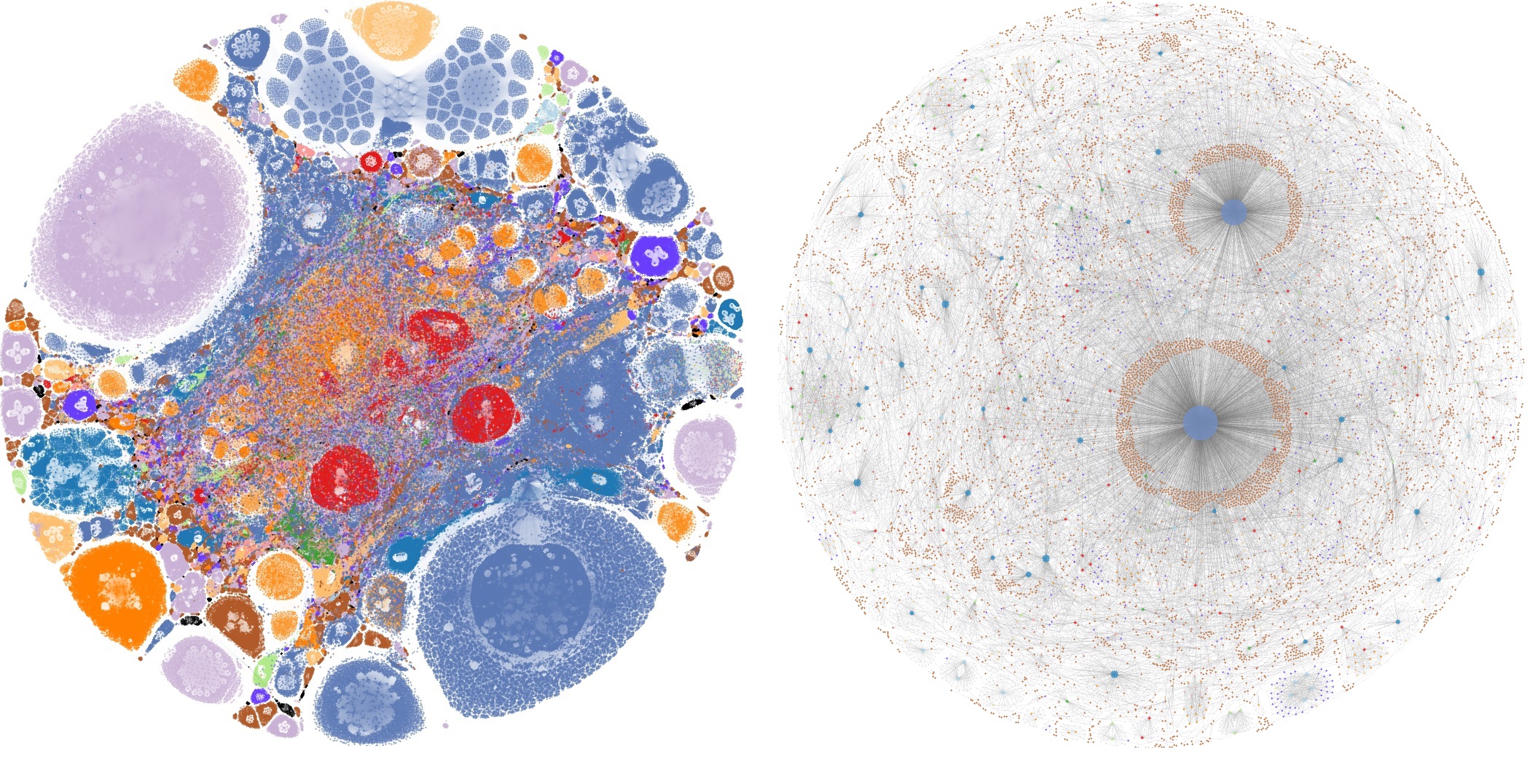}
 \caption{Effect of different values of degree thresholds on BerkStan: (two left) $\delta$, 
 and (two right) $5\delta$.}
 \label{fig:thresh}
\end{figure*}
\begin{figure}[h]
 \centering
 \includegraphics[scale=.1]{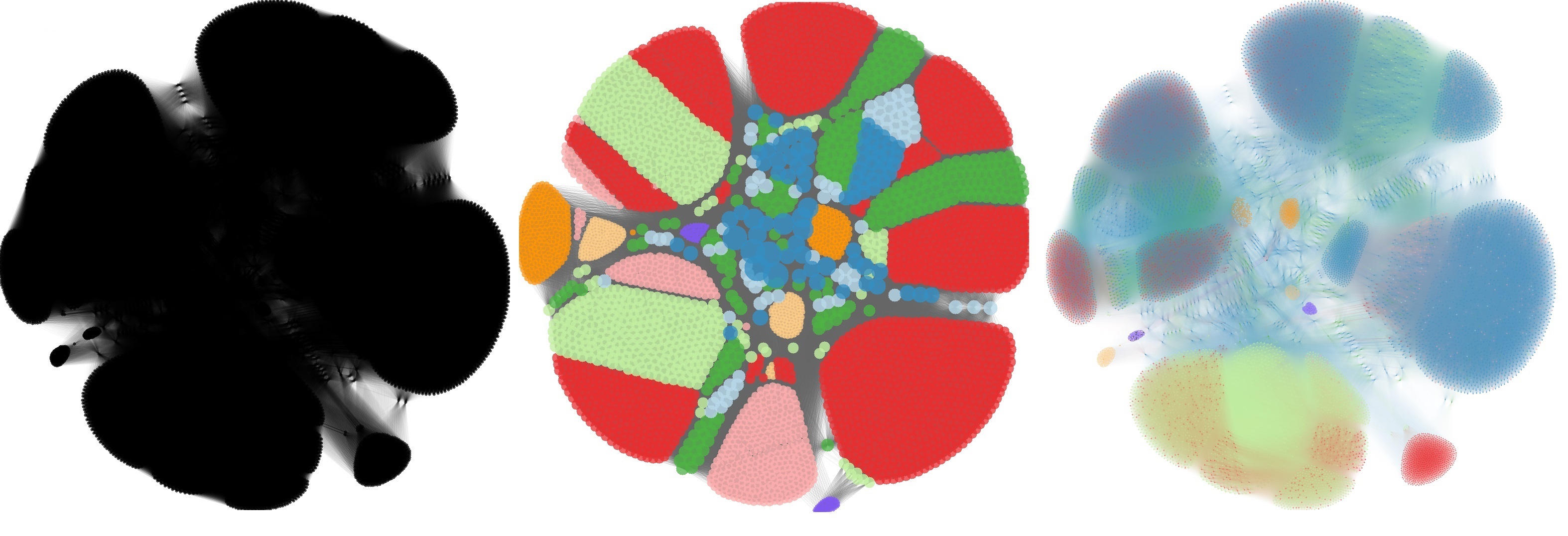}
 \caption{Visualization for the graph --- Authors.}
\end{figure}

\begin{figure*}[pt]
 \centering
 \includegraphics[scale=.08]{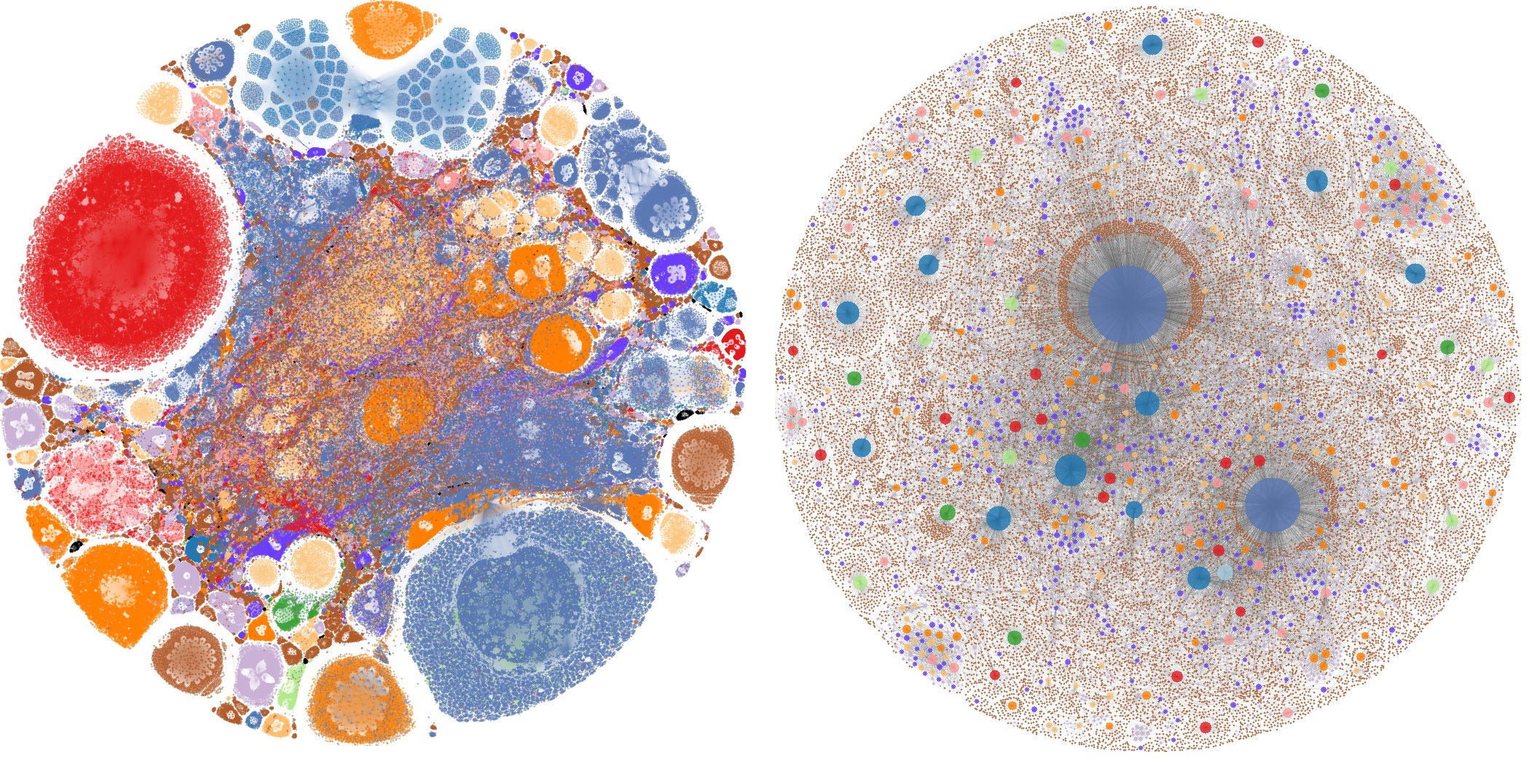}
 \includegraphics[scale=.08]{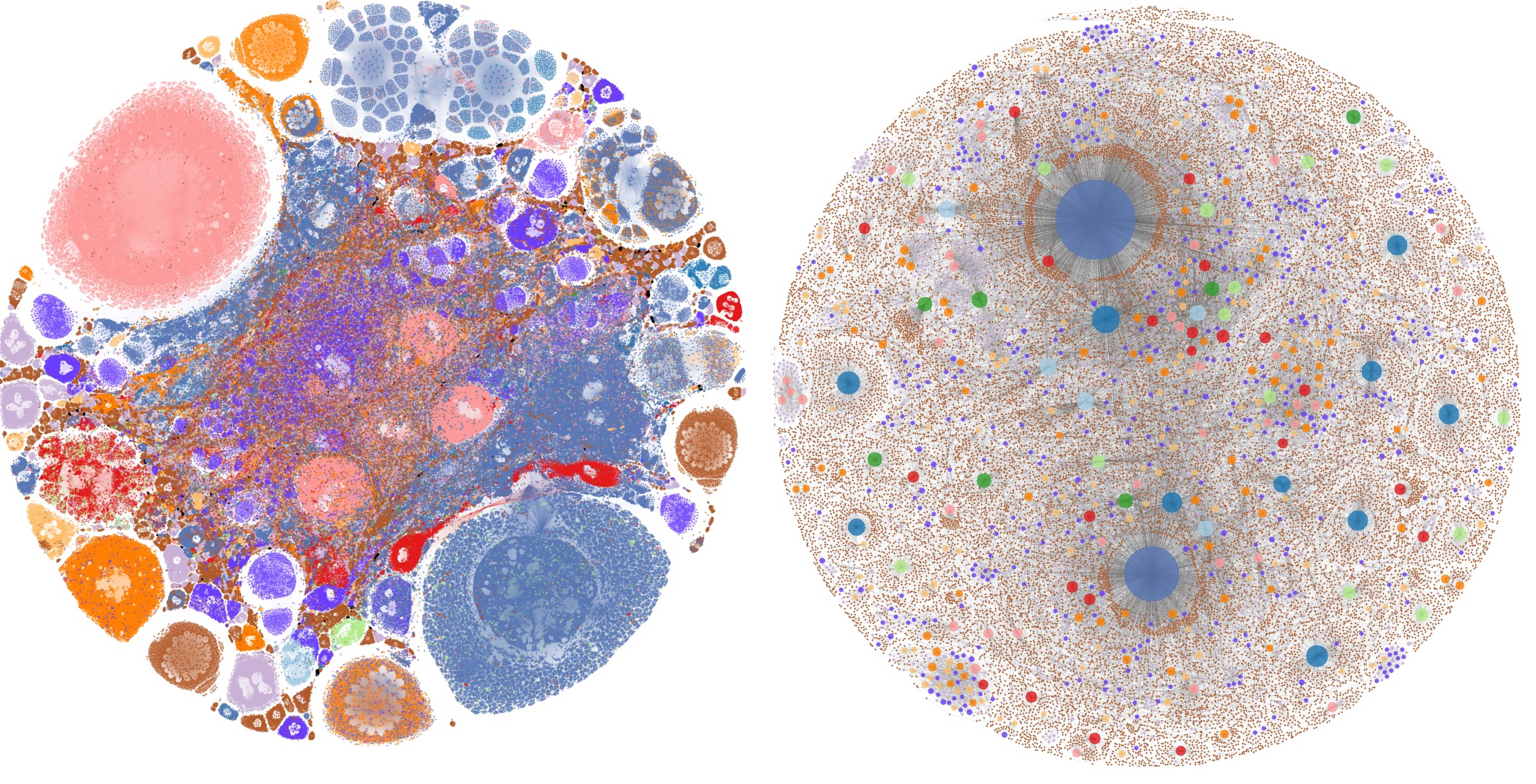}
 \caption{Effect of different sizes for count-min sketch matrix on web-BerkStan : (two left) 5K, 
 and (two right) 15K.}
 \label{fig:minsk}
\end{figure*} 
\begin{figure*}[pt]
 \centering 
   \includegraphics[scale=.06]{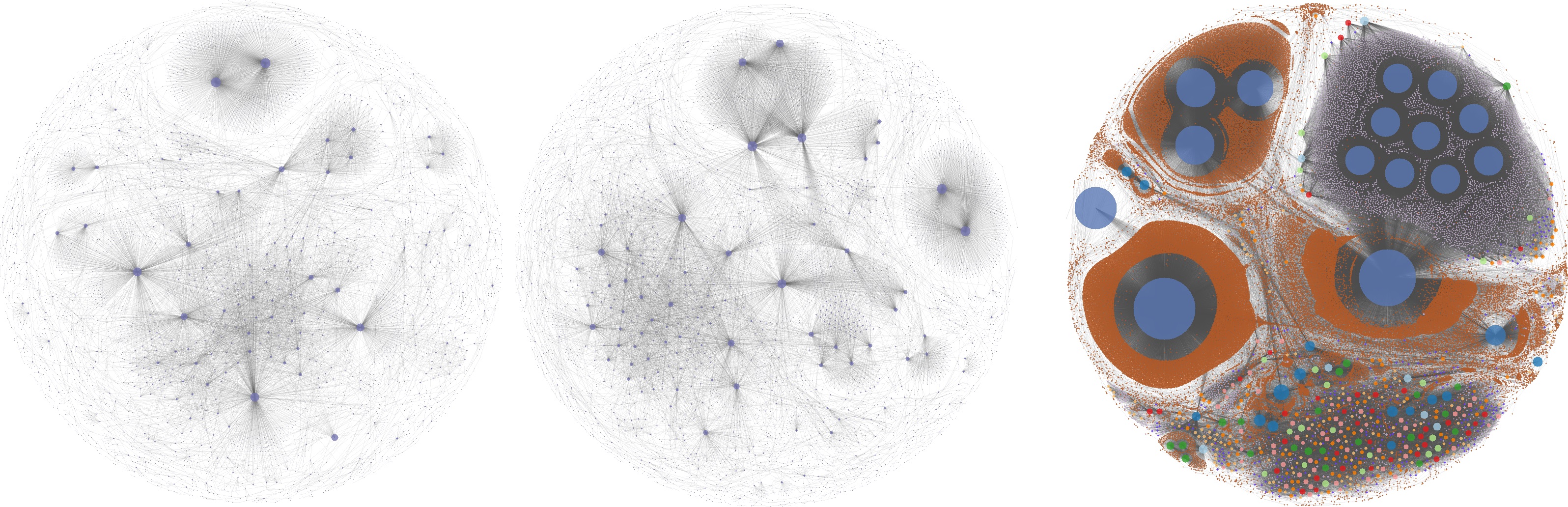}
   \includegraphics[scale=.06]{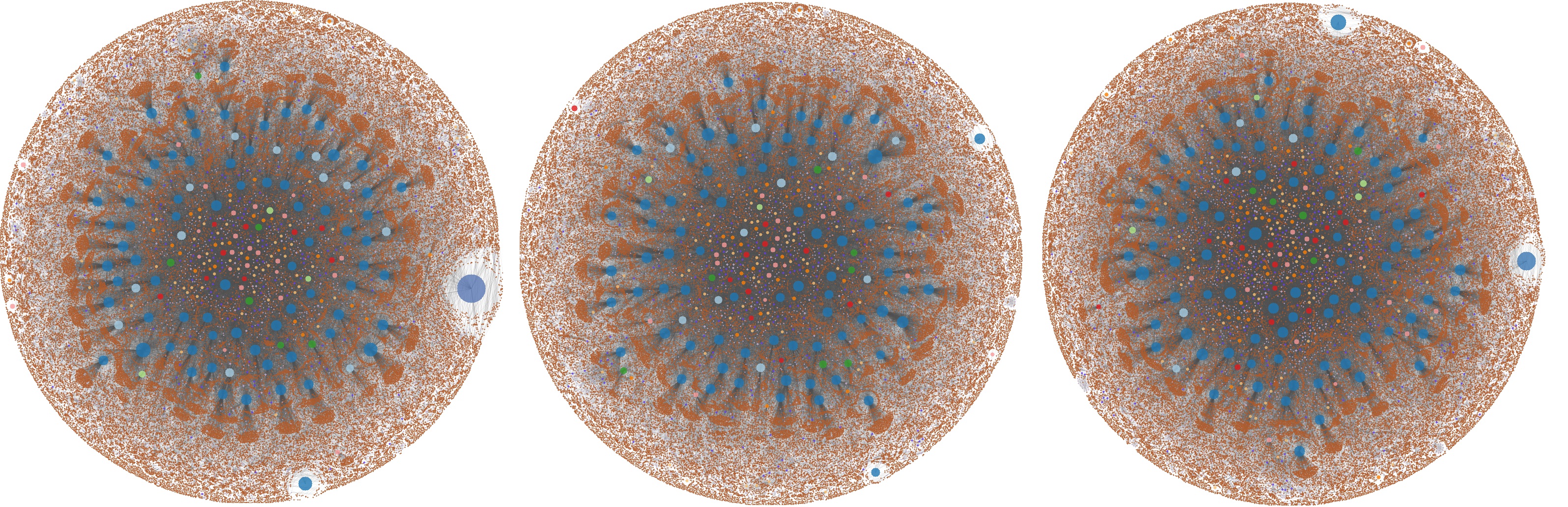} 
    \includegraphics[scale=.06]{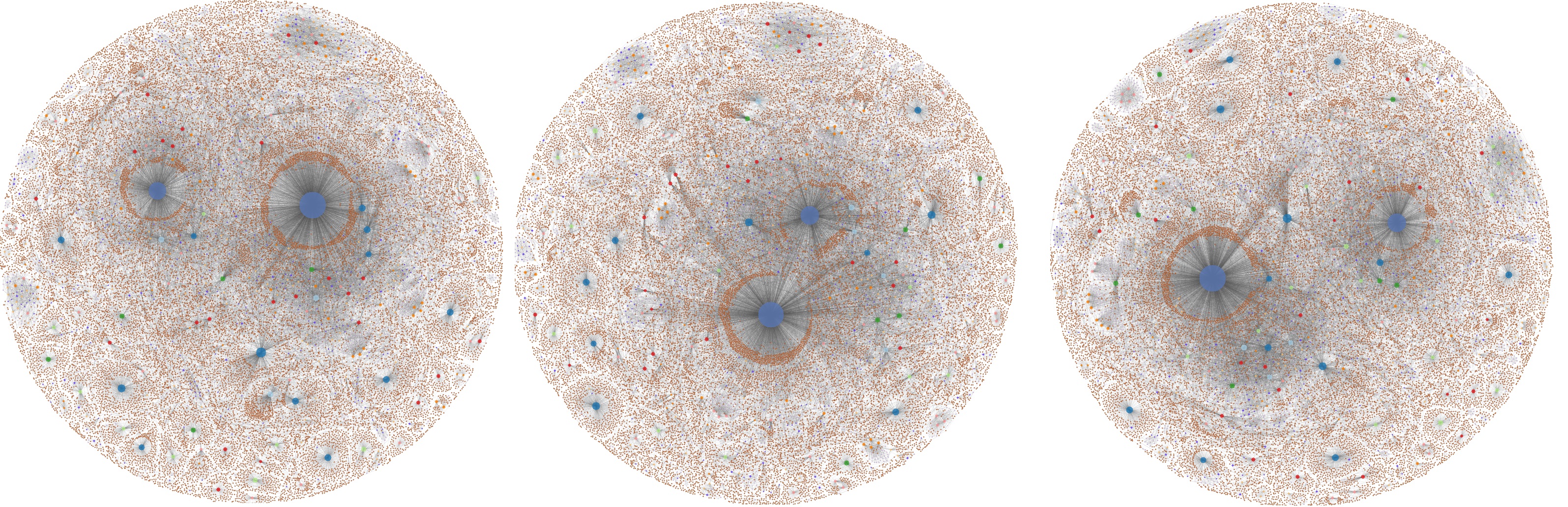} 
     \caption{Illustration for the effect of using (from left to right) two, three and four rounds for community detection for (from left to right 3 images for each graph) eu-2005, web-Google and web-BerkStan graphs.} 
     \label{fig:round}
\end{figure*}

\subsection{Data}
{We choose multiple real-world datasets for our work~\cite{leskovec2014snap,brinkmann2017exploiting,leskovec2009community,leskovec2010signed,leskovec2010predicting,leskovec2005graphs,mcauley2012image,yang2015defining,backstrom2006group}: Wiki-Talk (based on Wikipedia page edit), as-Skitter (based on Internet topology), web-flickr (based on shared images), github (based on software repositories), com-Youtube (based on user interaction), eu-2005 (based on web crawl), soc-LiveJournal (based on online interaction), web-Google and web-BrekStan (based on web page link). Whereas most of these graphs have millions of edges, they also have millions of nodes. Hence to examine dense graphs, we choose a graph Bio, created from bio-mouse-gene network~\cite{bansal2014community,nr}, and another graph called Authors. The Authors graph is created by taking authors of 15 journals as nodes, where an edge represents that the corresponding authors published in the same journal~\cite{tang2008arnetminer}.
}

\subsection{Results}
In this section we discuss the experimental results and how different parameter choices control the layout quality.
{
\subsubsection{Running Time} Table~\ref{tb-results} compares the running time of BigGraphVis (visualizing supergraph) and GPU-accelerated ForceAtlas2 (visualizing whole graph). 
For BigGraphVis, we report both the running time (in milliseconds) and 
the size of the supergraph (number of supernodes or communities detected), whereas for GPU-accelerated ForceAtlas2, we report the running time. We also compute the speedup in percentage for all the networks, which ranges between 70 to 95. The results are repeated for more than 20 times to see if there is any difference in speedup; and the least speedup is reported here.

Table~\ref{tb-results} also reports separately the time taken by BigGraphVis to detect the communities using 10 rounds. This is to provide an idea of time required to stylize a ForceAtlas2 visualization using a color mapping based on community sizes. We noticed that for all graphs this overhead is only a few seconds. Note that for all our graphs, the output were seen to converge in 3 rounds, which is an indication that the number of rounds could be lowered to achieve yet a smaller overhead.  
}

\subsubsection{Quantity measure (Modularity)} We examined  modularity of the detected communities using  \emph{modularity}~\cite{newman2006modularity}, which is a metric to measure the quality  of the communities. 
The modularity $Q$ ranges between $0$ and $1$, and computed using the following formula, where $A_{i,j}$ is the weight of the edge between vertices $i,j$, $k_i$ is the degree of node $i$, and $m$ is the total number of edges. The function $\sigma$ returns $1$ if $i$ and $j$ belong to the same community; otherwise, it returns $0$.  
\begin{equation}
Q=\frac{1}{2m}\sum_{i,j}^{v} [A_{i,j} - \frac{k_i k_j}{2m}]\sigma(C_i , C_j)
\end{equation} 
For five of the 10 graphs, the modularity scores were very high (above 0.7 and upto 0.9), and for none of them was below 0.55. This   indicates reliable detection of the communities. 

\subsubsection{Visual Comparison.} {Fig~\ref{fig:vis} illustrates three layouts for github, eu-2005, web-BerkStan and soc-LiveJournal graphs : (left) GPU-accelerated ForceAtlas2, (middle) BigGraphVis supergraph, and (right) ForceAtlas2 layout colored by BigGraphVis. It is noticeable that BigGraphVis were able to reveal big communities, although the speed up is gained by sacrificing the resolution that could be obtained from laying out the whole graph. Though we can access the members of each communities members using hierarchical community detection rounds created data set by BigGraphVis. For ForceAtlas2 layouts, which are colored by BigGraphVis, takes more time but shows a higher level of details. However, the community sizes seen in a ForceAtlas2 output may not always show their true sizes (i.e., the number of nodes or edges are not clear). On the other hand, the BigGraphVis supergraph can provide us with some idea of how big the big communities can be compared to the other communities in the graph and also provide a quick understanding of the number of big communities in a graph. }

Although true communities for these graphs are either unknown or not well-defined, for the graph Authors Fig~\ref{fig:around}, we know the authors are from 15 journals. Both the BigGraphVis supergraph and the ForceAtlas2 output colored by BigGraphVis, reveals about 15 big visual blobs. This provides an indication that even in cases when the streaming community detection may be a coarse approximation, BigGraphVis can produce a meaningful layout since it employs ForceAtlas2 to visualize the supergraph. 

 \subsubsection{Choices of Count-Min Sketch Parameters} {After community detection, BigGraphVis creates the graph summary by representing each community with a supernode. Each supernode is assigned a weight proportional to its number of edges, where the weight is computed using a count-min sketch data structure. The accuracy of the computed community sizes depends on the number of hash functions used in the count-min sketch.}

Table~\ref{tb-hash} illustrates the running time and communities detected for a different number of hash functions. Whereas the running time does increase significantly, the GPU memory required increases dramatically. Note that the number of communities detected often remains similar. However, as the number of hash functions increases, the sizes of the communities become more accurate based on the count min-sketch that provide more accuracy with more hash functions, which also improves the layout quality (e.g., see Fig~\ref{fig:hash}). The number of columns used in the count-min sketch matrix also has some impact on accuracy and thus on the layout (Fig.~\ref{fig:minsk}), but not as high as the number of hash functions.

\subsubsection{Choices of Community Detection Parameters} {BigGraphVis detects the community in several rounds, i.e., at each level, the current summary is further summarized hierarchically. However, the degree threshold (i.e., the mode degree $\delta$) used in the community detection algorithm restricts how big the communities can grow. Even when we choose the degree threshold $\delta^i$ for each round, we observe the summary (similarly, the layout) to becomes stable only after a few iterations when we inspect the layouts visually. }

{
Table~\ref{tb-iteration} illustrates the running time (in millisec), number of communities and their sizes for two, three and four rounds. Note that each round is a complete independent run of the algorithm and thus in the table, the number of communities may not always be decreasing. However, as expected, the number of inner edges can be seen to increase with the number of rounds. In fact the intra-community edges per community also grow with the number of rounds, which cannot be seen from the table, but from some visual inspection (Fig.~\ref{fig:round}).
}

\begin{figure}[h]
 \centering
   \includegraphics[scale=.07]{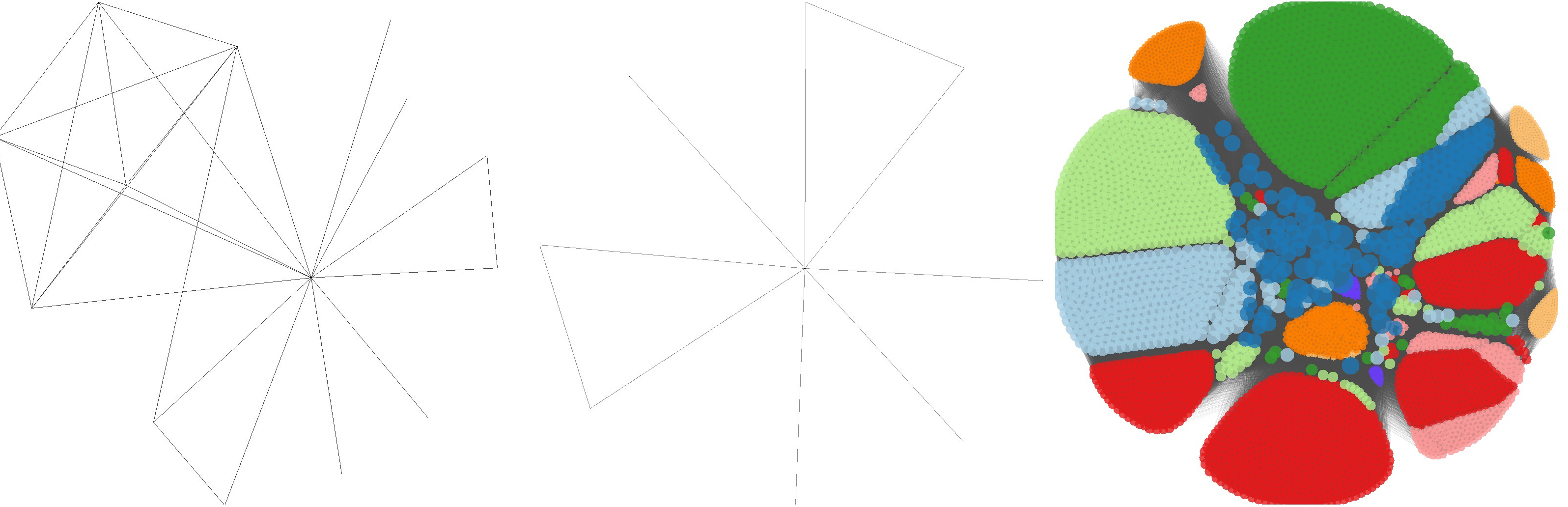}
     \caption{Illustration for the effect of different rounds of community detection for the graph --- Authors.}
\label{fig:around}
\end{figure}

Although some graphs show high similarity in layout for different numbers of rounds, for others, e.g., bio-mouse-gene, as-Skitter, and eu-2005, the growth in community sizes are noticeable. The case for the graph Authors is particularly interesting. Since the mode degree is 2, it is expected that to form big communities, it would require multiple rounds, and the transition is clearly noticeable over the increasing number of rounds (Fig.~\ref{fig:around}). In addition, from Table~\ref{tb-iteration}, we can observe that for this graph in round four, the number of inner-community edges highly increased, which inherently indicates that the intra-community edges should also increase. This can also be seen visually from the layout.

{We did not find the degree threshold to be a very sensitive parameter (when inspected visually), as long as it is closer to the mode degree $\delta$. For example, Fig.~\ref{fig:thresh} illustrates the effect of choosing $\delta, 3\delta$ and $5\delta$ as the degree thresholds.}

 \begin{figure*}[h]
 \centering
 \includegraphics[width=\textwidth]{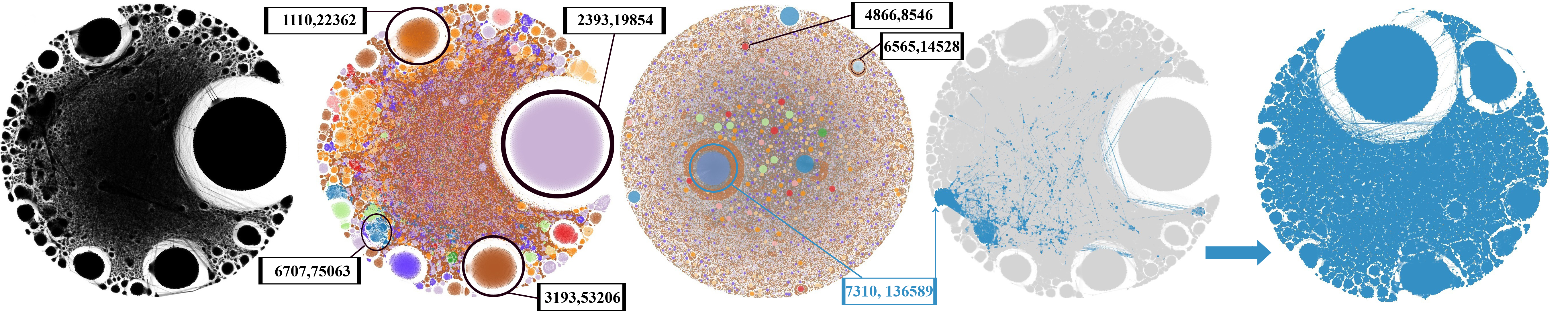}
  \includegraphics[width=\textwidth]{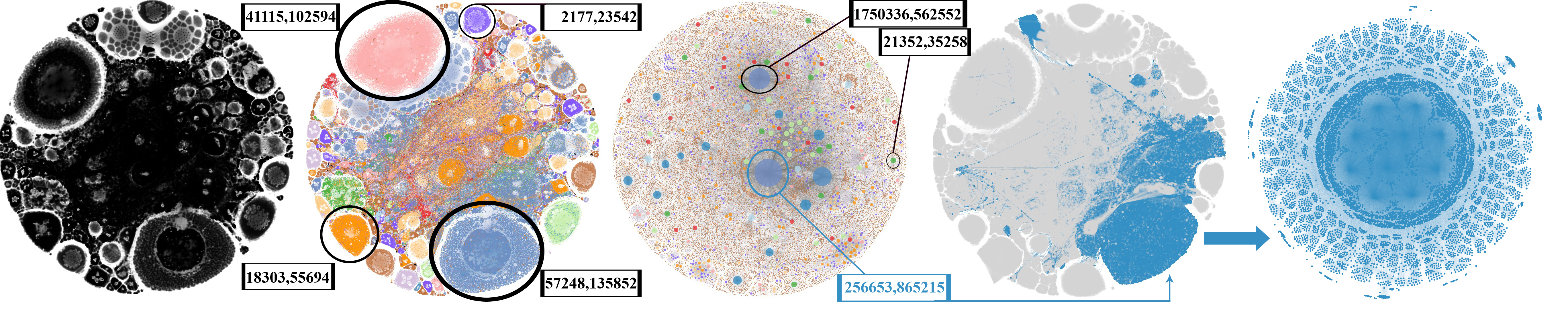}
  \includegraphics[width=\textwidth]{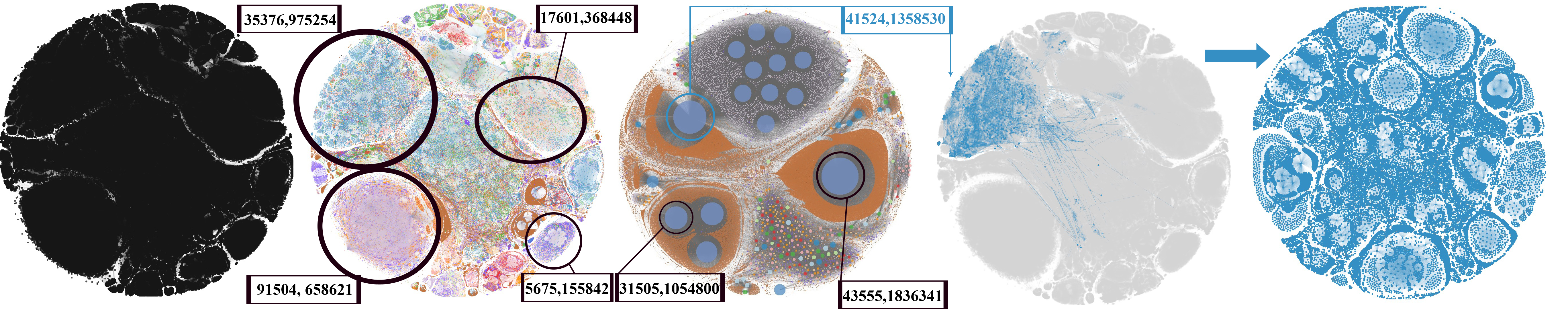}
 \caption{Analysis of clusters and supper nodes for ForceAtlas2 and BigGraphVis visualizations. (top) github, (middle) web-BerkStan and (bottom) eu-2005.}
 \label{fig:comp}
 \end{figure*}
  \begin{figure*}
 \centering
 \includegraphics[scale=.09]{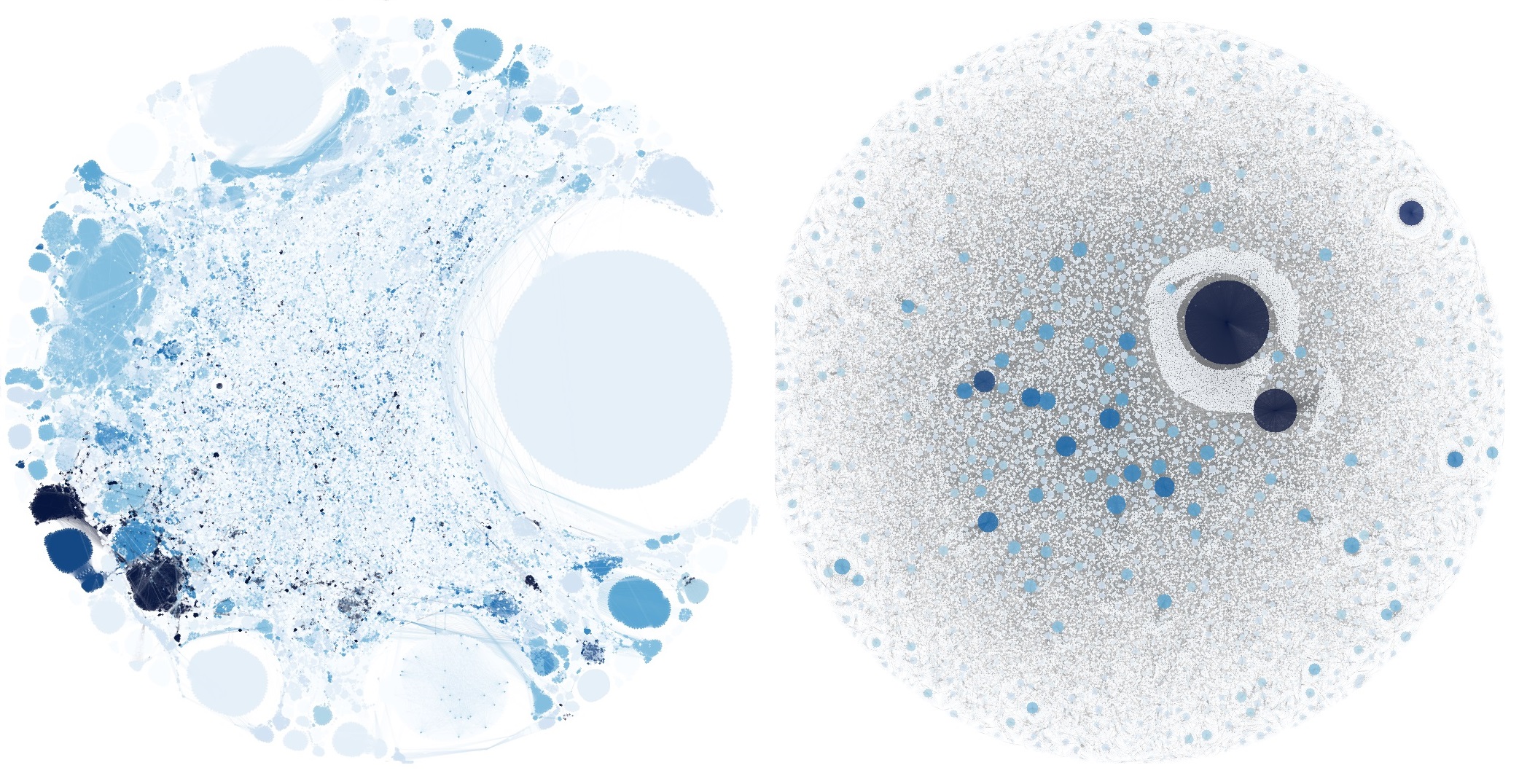}
  \includegraphics[scale=.09]{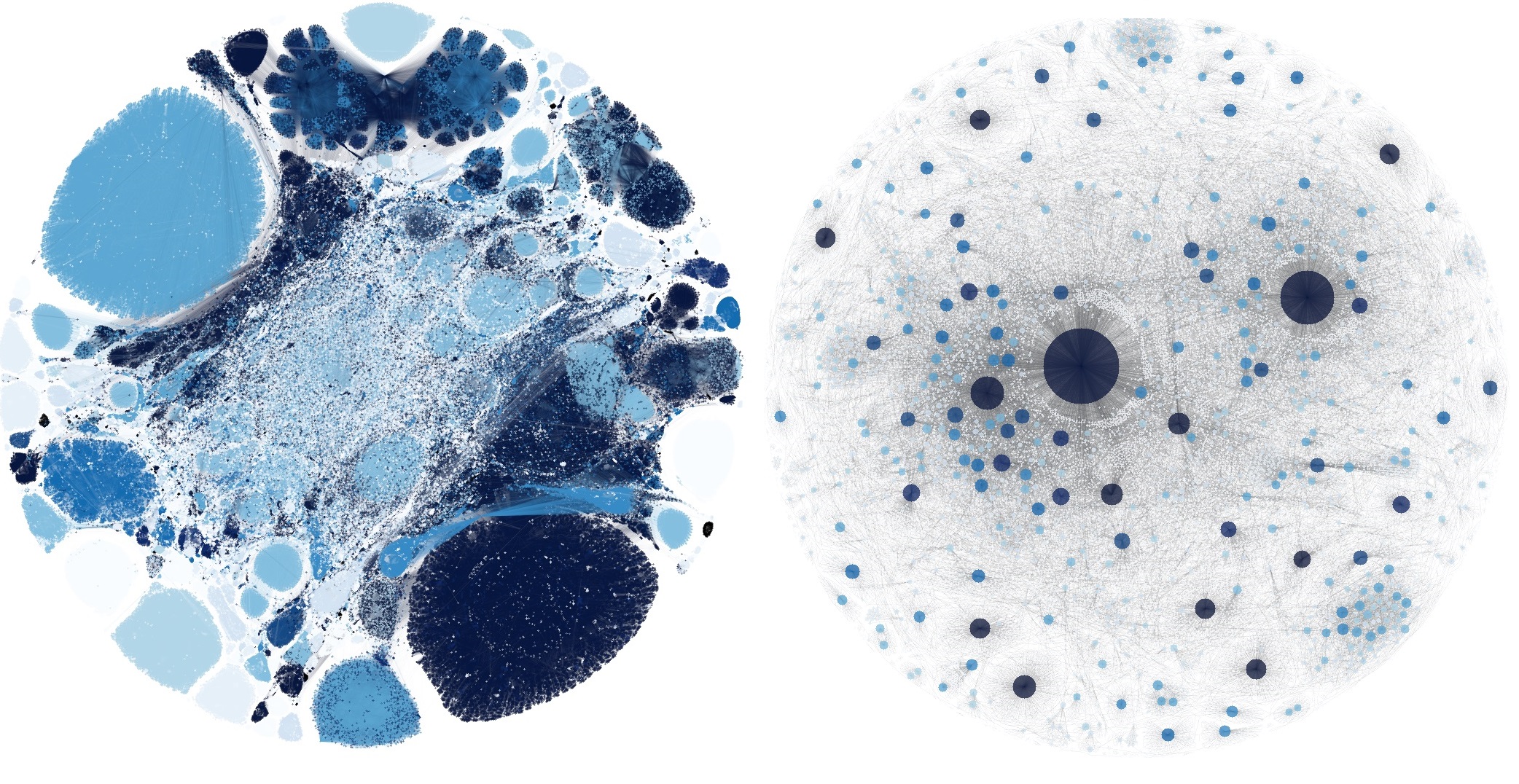}
  \includegraphics[scale=.09]{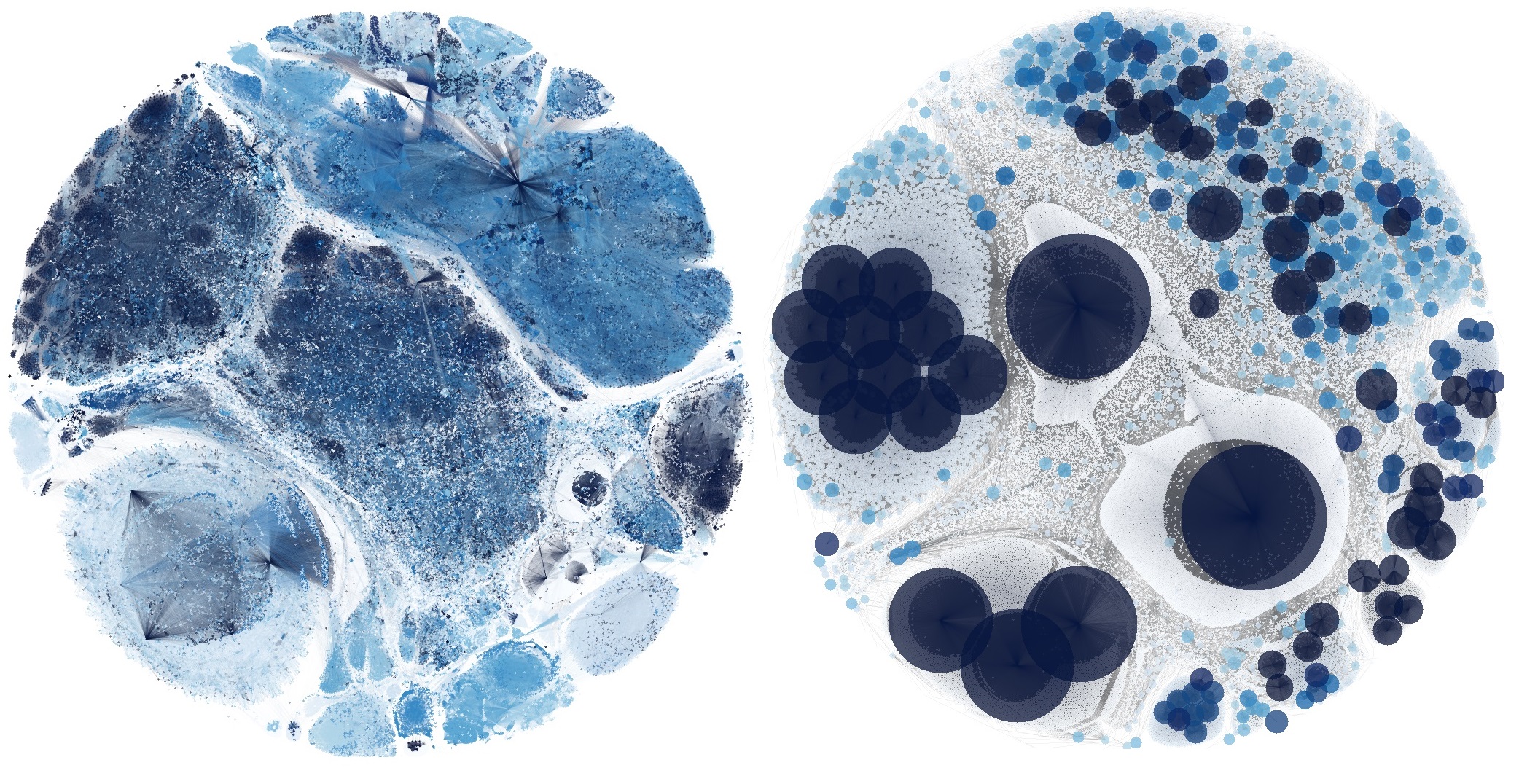}
 \caption{ForceAtlas2 and BigGraphVis visualizations for github, web-BerkStan and eu-2005 using sequential colouring. Two images per dataset. }
 \label{seq}
 \end{figure*}

\section{Case Study}
To investigate our method's reliability,  we study the visualization of the github, web-BerkStan and eu-2005 datasets. Figure~\ref{fig:comp} shows the visualization for these graphs in 3 rows.  
 Each row has five images: (from left) first a visualization of the entire graph using ForceAtlas2, 
 a colored ForceAtlas2 with four labeled clusters representing clusters number of nodes and edges, 
 BigGraphViz with three labeled clusters, revealed community from the largest  BigGraphViz supernode, 
 and finally, ForceAtlas2 visualization of the selected supernode. 
 The labeling is not in any specific order. 
 While labeling the clusters, we choose a  cluster or suppernode that appears to occupy a large circular area, and then choose a few  other clusters of different sizes. 
 
\paragraph{\bf Cluster Size Interpretation  in ForceAtlas2 Visualization:} 
We can observe several instances in  ForceAtlas2 visualizations, were the relative sizes of the clusters are counterintuitive. For example, the largest-area cluster (purple) in github visualization (Figure~\ref{fig:comp}(top)) appears to have only 2k vertices and 19k edges, whereas a cluster (blue) with half the area contains 6k vertices and 6k edges. 
Similarly, in the ForceAtlas2 visualization of web-BerkStan (Figure~\ref{fig:comp}(middle)), the purple cluster with only 2k vertices and 23k edges appears to be similar to a much larger orange cluster (18k vertices, 55k edges). For
 eu-2005 dataset (Figure~\ref{fig:comp}(bottom)), we can again see two clusters with similar area with a very different combination of vertices and edges: one with 35k vertices and 975k edges, and the other  with 91k vertices and 658k edges.

\paragraph{\bf Locating BigGraphVis Supernodes in ForceAtlas2 Visualization:} The supernode sizes in BigGraphVis visualization are relative to the number of edges in the corresponding community, the interpretation of the node sizes appears to consistent.  Therefore, we were more interested to examine  the location of communities detected by  BigGraphVis in the ForceAtlas2 visualization.  While the qualitative color coding provides us with some idea, for a better understanding, we took looked into it from two different perspectives. 

The first one is coloring  BigGraphVis communities with a sequential color coding (Figure~\ref{seq}). From the sequential coloring, we can observe that the large supernodes detected by the BigGraphVis (darker color) do not always identify the large-area cluster in the ForceAtlas2 visualization. Although for the eu-2005 dataset and web-BerkStan datasets  we can see the large supernodes to span some large-area clusters, there are some large-area clusters that in ForceAtlas2 visualization that are not identified as large community in BigGraphVis. This can clearly be seen the github dataset.

\paragraph{\bf Reliability of Supernodes in BigGraphVis  Visualization:} 
Although we expect BigGraphVis to miss some communities as it trades off some quality to achieve considerable speed, we were interested to see whether we could rely on the communities detected by the ForceAtlas2. We observed that the detected communities indeed provides reasonable modularity score (Table~\ref{tb-results}). Furthermore, when we extract and visualize the community separately using ForceAtlas2, we obtain a visualization showing reliable community structure. The  fourth column of Figure~\ref{fig:comp} illustrates the location of a large community detected by BigGraphVis in the ForceAtlas2 visualization. The fifth column illustrates the community structure revealed by ForceAtlas2, providing an evidence that the community detected is reliable; although it may not always create a large-area cluster in the ForceAtlas2 visualization.

\section{Limitations}
Since ours is a foundational work that brings streaming community detection into GPU-accelerated graph visualization, we invested our effort entirely on the BigGraphVis's reliability by investigating on parameter sensitivity and community mapping between BigGraphVis and ForceAtlas2 visualizations. A comparison with a variety of other existing big graph visualization approaches can be the obvious next step of this work. Another interesting avenue for  exploration would be to interpret the reliability of the distances between the clusters and supernodes in the ForceAtlas2 and the BigGraphVis visualization.  

We used visual analysis for a few parameter choices, and  for the comparison between ForceAtlas2 and BigGraphVis visualizations.  Since the visual analysis is subjective,  it would be interesting to conduct a formal user study to investigate real-life use cases and tasks.

\section{Conclusion}
{
In this paper, we propose BigGraphVis that  visualizes graph summaries based on community detection. We show how to leverage streaming community detection and GPU computing to further improve the cutting edge approaches to graph visualization. Through a detailed experiment with the real-world graphs (the biggest graph, soc-LiveJournal, had about 34 million edges), we observed that BigGraphVis can produce a meaningful summary within a few minutes (about five minutes for soc-LiveJournal). We also observe that the graph summary produced by BigGraphVis can be used to color ForceAtlas2 output to reveal meaningful graph structure. However, in addition to ForceAtlas2 parameters, BigGraphVis uses some additional parameters for streaming community detection. Although we examined the effect of the parameters and suggested default values, future research on improving this approach further by removing such dependencies on parameters would be interesting. We believe that our work will inspire future research on leveraging streaming algorithms and GPU computing to visualize massive graphs.
}

\section*{Acknowledgment}
This work is supported by the Natural Sciences and Engineering Research Council of Canada (NSERC), and by two Canada First Research Excellence Fund (CFREF) grants coordinated by Global Institute for Food Security (GIFS) and Global Institute for Water Security (GIWS).

\bibliographystyle{abbrv}

\bibliography{arxivtemplate}
\vfill
\end{document}